\documentclass[aps,tightenlines,longbibliography,notitlepage,nofootinbib]{revtex4-2}

\usepackage{amsmath}
\usepackage{amssymb}
\usepackage{bbm}
\usepackage{simplewick}
\usepackage{mathtools}
\usepackage[normalem]{ulem}
\usepackage{dsfont}
\usepackage{mathrsfs}
\usepackage[makeroom]{cancel}
\usepackage{graphicx}
\usepackage{bm}

\def\rhat{{\hat {\boldsymbol r}}}
\def\khat{{\hat {\boldsymbol k}}}

\DeclareMathOperator\erf{erf}
\DeclareMathOperator\sign{sgn}

\def\0{\boldsymbol{0}}
\def\RR{\mathbbm{R}}

\def\BD#1{{\boldsymbol{{#1}}}}

\def\Lag{{\cal L}}
\def\ii{\mathrm{i}}

\def\bnabla{{\boldsymbol\nabla}}

\def\OM#1{\omega_{#1}}

\def\B#1{\left(#1\right)}
\def\BB#1{\left[#1\right]}

\def\la{\langle}
\def\ra{\rangle}

\def\lara#1{\la#1\ra}

\def\be{\begin{equation}}
\def\ee{\end{equation}}

\def\hc{\text{h.c.}}
\def\cc{\text{c.c.}}

\def\for{\ \text{for} \ }

\usepackage[ddmmyyyy]{datetime}
\makeatletter
\def\Dated@name{}
\makeatother

\def\pref{0.5}
\def\cl{\text{cl}}

\def\jmat{j_\text{mat}}
\def\KS{{\BD{k}\sigma}}

\begin{document}
\title{Expanding Coulomb field}
\author{Bogdan Damski}
\affiliation{\text{Jagiellonian University,
Faculty of Physics, Astronomy and Applied Computer Science,}
\\ {\L}ojasiewicza 11, 30-348 Krak\'ow, Poland}
\begin{abstract}
We briefly  discuss vector field theories that allow for electric charge non-conservation
and  focus on the modified version of the classical Maxwell theory,
where the Lagrangian density  has
the same structure as in the Gupta–Bleuler theory  with 
an  external  current--vector field coupling.
Within this framework, 
we study the causal dynamics of the field configuration 
following the sudden creation of an electric charge modeled by an external current.
The resulting solution describes an expanding  Coulomb field, which is an
inherently dynamical object featuring a shockwave-like front propagating 
at the speed of light 
(the Coulomb field is confined to the volume swept out by this front).
We obtain intriguing results for the energy of
the studied field configuration and for the energy carried by the shockwave-like front.
General results on the dynamics following the sudden onset of an
otherwise static external current are given.
The field equations are solved in an unconventional way by means of 
a quantum-to-classical mapping. An energy-based 
argument against charge non-conservation in the Maxwell theory is 
presented.
\end{abstract}
\date{\today}
\maketitle

\section{Introduction}

Static fields require a field source that has existed for all 
time \cite{GriffithsBook,MaggioreBook}. 
However, this condition is never met by laboratory-created
sources and is questionable
if one assumes that  the age of the Universe is finite.
Consequently, expanding fields, 
which causally transmit information about their source, 
are more fundamental than their static counterparts. 
These considerations motivate our comprehensive studies of expanding fields, 
which began with \cite{BDmag,BDProca} and
are  continued in this work.

The non-equilibrium quantum dynamics of a vector field,
following a sudden switch-on of an otherwise static external current,
was  discussed in \cite{BDmag,BDProca}.
The problem was analyzed in two settings.
The massless vector field, described by the Maxwell theory, was investigated
in \cite{BDmag}, where the switched-on current has vanishing charge density
and non-zero charge current. This leads to the formation  of a shockwave-like
front propagating at the speed of light. In the region swept by it,
a static magnetic field is established, whereas in the region not 
yet reached by it, the field is absent for causal reasons.
The massive vector field, described by the Proca theory, was studied 
in \cite{BDProca}. In this case, the switched-on current has non-zero
charge density and vanishing charge current. This produces  the dynamics 
similar to the above described except a static timelike rather than 
spacelike component of a vector field 
is established within the region swept by a propagating front,
and this occurs on a long timescale compared to the massless case.
These remarks indicate that at any given time the discussed field configurations 
have an {\it expanding} rather than static character.

Encouraged by the findings of \cite{BDmag,BDProca}, we would like to 
study  the classical  expanding  Coulomb field. This amounts to  discussing 
the field configuration resulting from the creation of an electric charge.
The problem now is that such a  process is forbidden in the Maxwell
theory (see e.g. Sec.~IIC of \cite{Nieto_RMP2010}).
In the  search for physical phenomena beyond the current status quo, 
one may  ask how  the Maxwell theory could be modified to 
accommodate electric charge non-conservation.
We propose such a modification  in this work
and show that it leads to 
a natural causal picture for the dynamical generation of the Coulomb field.
Several remarks are in order now.

First, the possibility of electric charge non-conservation has been studied
for many decades and remains an interesting subject of research. Reviews of the early
investigations can be found in \cite{Okun1989} and Sec.~IIC of \cite{Nieto_RMP2010}. 
Recent experimental efforts are discussed in \cite{ArnquistNatPhys2024,ZhangPRD2025,AgostiniPRL2015}.

Second, as we employ the external current approach, we do not propose any mechanism for electric
charge non-conservation. Note that the external current is a prescribed source.

Third, electric charge conservation in the  Maxwell theory
can be seen as a consequence of relativistic
invariance and  masslessness of a  vector field
(see e.g. \cite{Weinberg1964}).
Thus, it is unsurprising that a causality-related
problem arises in \cite{BermanAmJPhys2008}, where
the issue of the dynamical creation of a Coulomb
field was investigated within
the Maxwell theory (see Appendix \ref{Difficulties_app}
for  details).
In this context, we would like to underscore that
relativistic invariance and  masslessness of a vector
field are kept intact in our work.
A brief discussion of the expanding Coulomb field can be
found in \cite{AlickiJPA1978}, a related work we encountered
only when our manuscript was already nearly finished.
Our work represents both
an extension of \cite{AlickiJPA1978} and a continuation
of \cite{BDmag,BDProca} for the following
reasons.
Our construction of the model allowing for electric charge
non-conservation differs from that in \cite{AlickiJPA1978}, yet 
both approaches remarkably lead to the same model.
Our solution technique and the organization of our discussion,
which closely follow the spirit of \cite{BDmag,BDProca},
contrast with \cite{AlickiJPA1978}.
Finally,
we consider the creation of an extended charge distribution. 
This leads to well-defined expressions, in contrast to the formal ones obtained 
in \cite{AlickiJPA1978}, which is particularly important for understanding the energy of the expanding Coulomb field.

The outline of this paper is the following. 
After considering some options,
a modification of the  Maxwell theory allowing  for
electric charge non-conservation is proposed  in Sec. \ref{Basic_sec}.
The basic properties of the resulting theory as well as 
a clear statement of the problem we aim to solve  
are presented in Sec. \ref{Classical_sec}.
The general results concerning the classical
dynamics following the sudden switch-on of an 
external current are derived in Sec. \ref{Quantum_sec}. 
In this section, instead of solving classical differential
equations, we quantize the classical theory, solve the time-dependent Schr\"odinger equation,
and associate  the classical quantities  of interest with the expectation values of
the relevant operators (the energy- and vector field-related quantities are
obtained in this way).
A specific form of the suddenly switched-on external current, 
describing a Gaussian-shaped charge distribution at rest,
is assumed in Sec. \ref{Expanding_sec}. In this section,  
taking advantage of the exact solvability of the problem,
we present the detailed discussion of the expanding Coulomb field.
Our results are summarized  in Sec. \ref{Discussion_sec},
where besides a brief recap,  we compare 
the quantum theory employed in Sec. \ref{Quantum_sec} 
with  the Gupta-Bleuler  theory describing a covariantly quantized 
massless vector field.


\section{Basic insights}
\label{Basic_sec}

Electric charge conservation in the Maxwell theory can be seen as the 
consequence of the structure of field equations
\be
\partial_\mu F^{\mu\nu}_\cl=\jmat^\nu,
\label{Maxx}
\ee
where 
$F^\cl_{\mu\nu}\equiv\partial_\mu A^\cl_\nu-\partial_\nu
A^\cl_\mu$, $A_\cl$ denotes the vector field,  
 $\jmat$ stands for the external current,
and $\cl$ stands for classical (see Appendix \ref{Conventions_sec}
for our conventions).
Namely, acting with $\partial_\nu$ on (\ref{Maxx}), one obtains 
$\partial\cdot\jmat=0$.\footnote{As is well known, 
this equation does not imply 
electric charge conservation
when $\BD{j}_\text{mat}$  is not sufficiently
spatially localized (see   Appendix \ref{Difficulties_app} for an  example). 
We assume in this work that $\int d^3r \jmat^0$ is well defined  and that 
$\lim_{r\to\infty}r^2\int d\Omega(\rhat)\rhat\cdot\BD{j}_\text{mat}=0$, where $\Omega(\rhat)$
is  a solid angle in real space.}
To explain the strategy for finding a reasonable 
modification of the Maxwell theory breaking the  conservation law, 
we consider the  following  field 
equations\footnote{See \cite{MullerPRD2005} and references therein for a 
discussion of Lorentz symmetry-violating models allowing 
for electric  charge non-conservation (such models are not considered in this work).}
\be
\partial_\mu F^{\mu\nu}_\cl=\jmat^\nu -\Phi_\cl^\nu.
\label{MAXX}
\ee
Acting with $\partial_\nu$ on  (\ref{MAXX}),
we arrive at $\partial\cdot\jmat=\partial\cdot\Phi_\cl$,
which  implies  that $\partial\cdot\jmat$ need not vanish.
This property  allows for describing field configurations resulting from 
the process of   electric charge creation.

Next, we observe that there is no unique way of extending the Maxwell 
theory to allow for electric charge non-conservation. In other words,
there is no unique way of choosing the field $\Phi_\cl^\mu$.
To discuss this issue, we consider constructing $\Phi^\mu_\cl$ out of the vector field itself.
For example, one may take into account
(i) $\Phi_\cl^\mu\propto A^\mu_\cl$,
(ii) $\Phi_\cl^\mu\propto \Box A^\mu_\cl$, and 
(iii) $\Phi_\cl^\mu\propto \partial^\mu(\partial\cdot A_\cl)$.
These options preserve translational invariance (no explicit dependence on spacetime
coordinates), the second-order differential character of  field 
equations, and linearity of the vector-field sector of the theory. Thereby,
they satisfy reasonable theoretical requirements. 
Their  physical implications substantially  differ.

Option (i) leads to the Proca theory describing massive vector bosons, provided
that the sign of the  coupling constant is appropriately chosen.
This theory   was investigated in the charge-creation context in
\cite{BDProca}.
Option (ii), in the presence of a conserved  external current,
yields solutions that differ from those of  
the Maxwell theory. In a sense, the
correspondence principle is violated if the goal of introducing 
$\Phi_\cl^\mu$ is merely to allow electric charge non-conservation;
the same issue appears in option (i).
Option (iii)  yields a theory whose solutions reproduce those of
the Maxwell theory  when an  external current is conserved, provided that 
both $\partial\cdot A_\cl$ and $\partial_t(\partial\cdot A_\cl)$ vanish at some
initial time,  which is hardly an unreasonable requirement.

These observations leave us with   option (iii). 
Specifically, we propose 
replacing $\Phi_\cl^\mu$ in the field equations  by 
$\partial^\mu(\partial\cdot A_\cl)$, where the unit prefactor is chosen so that the field equations 
of the resulting theory superficially take the Lorenz-gauge form.
The desired modification of the Maxwell theory is then enforced
by the following modification of the Lagrangian density
\be
-\frac{1}{2} (\partial\cdot A_\cl)^2,
\label{wdek}
\ee
leading to the theory discussed in the subsequent  
sections.

Finally, we note that (\ref{wdek}) is known as the gauge-fixing term
in the covariant quantization scheme of a massless vector field \cite{Greiner}.
In this scheme,  (\ref{wdek})   is incorporated  into the Lagrangian density to resolve 
the  well-known problem of the Maxwell theory: the absence of the
momentum canonically conjugate to the timelike component of the vector field 
(see \cite{GieresNPB2021}   for a recent overview). 
Remarkably, in our case, it is introduced for an entirely different purpose: 
to enable the response of a vector field to electric charge creation.
Moreover, the same modification arises in a certain extension of the 
quaternionic formulation of the Maxwell theory
(see \cite{AlickiJPA1978} and the discussion therein).
Finally, the same modification of the Lagrangian density also appears 
in classical considerations that go beyond the  Maxwell theory
(see  e.g. \cite{ModaneseSym2023} and references therein as well as
\cite{JimenezPRD2011} for a  discussion in a cosmological context).


\section{Model  and problem setup}
\label{Classical_sec}

We consider the following Lagrangian density
\begin{subequations}
\begin{align}
\label{LG1}
{\cal L}_\cl&= -\frac{1}{2} (\partial_\mu A^\cl_\nu)^2 - \jmat^\mu A^\cl_\mu\\
&=-\frac{1}{4} (F^\cl_{\mu\nu})^2 -\jmat^\mu A^\cl_\mu -\frac{1}{2} (\partial\cdot A_\cl)^2
+ \text{4-divergence}.
\end{align}
\label{LG}%
\end{subequations}
In the following, we discuss basic properties of theory (\ref{LG}), the ones
relevant for our subsequent studies.

First, the field equations are
\begin{align}
\label{elf}
&\partial_\mu F_\cl^{\mu\nu}= \jmat^\nu + j^\nu_\text{int},\\
&j^\nu_\text{int}= - \partial^\nu(\partial\cdot A_\cl),
\end{align}
which implies that $\jmat^\mu +j^\mu_\text{int}$ 
plays the role of the conserved current. 
This  current is composed
of the  matter field-based  external current, $\jmat^\mu$,  and 
the vector field-based  {\it internal} current, $j^\mu_\text{int}$.
The conservation of $\jmat^\mu +j^\mu_\text{int}$  implies  
\be
\partial\cdot\jmat= \Box(\partial\cdot A_\cl),
\label{Cdyn}
\ee
explicitly showing that the electric   charge, i.e. the charge  associated 
with the external current, does not have to be conserved.

Second, in the Maxwell theory, one necessarily deals with
electric charge conservation. On the technical level, 
this follows from field equations lacking an internal-current contribution. 
On the conceptual level, 
one  explains this fact via a simple argument 
\cite{Nieto_RMP2010}.\footnote{See Sec. \ref{Energy_sub} for an energy-based
argument against the possibility of electric charge non-conservation in the 
Maxwell theory.}
Namely, one uses   Gauss' law to argue that 
the modification of the total electric charge implies instantaneous 
modification of the electric field at arbitrarily large distances,
which is prohibited by special relativity. This observation 
reflects the non-local implications of  Gauss' law.
We would like to underscore that 
theory (\ref{LG})  is relativistic and it does
offer the
possibility of electric charge non-conservation.
The above argument fails 
for theory (\ref{LG}) because one deals in it  with modified Gauss' law
(\ref{qqqq1})
due to the presence of the internal current.

Third, field equations  (\ref{elf})  can be written as
\be
\Box A^\mu_\cl= \jmat^\mu,
\label{boxAj}
\ee
which look like the Lorenz-gauge field equations in the 
Maxwell theory. However, note that there is an implicit assumption in the
Maxwell theory that $\partial\cdot\jmat=0$, which is not present in 
theory (\ref{LG}).

Fourth, there is another  theory of the vector field featuring
an internal current. Namely,  the Proca theory of a  massive
vector field, say  $V_\cl$, which is introduced via
\be
{\cal L}^\cl_\text{Proca}= -\frac{1}{4}
(\partial_\mu V^\cl_\nu-\partial_\nu V^\cl_\mu)^2
+\frac{m^2}{2} (V^\cl_\mu)^2
-\jmat^\mu V^\cl_\mu,
\label{ProcaL}
\ee
where $m$ is the mass of the vector boson. In this theory,
the  field equations read  
 $\partial_\mu (\partial^\mu V_\cl^\nu-\partial^\nu V_\cl^\mu)
 =\jmat^\nu -m^2 V_\cl^\nu$. Hence, 
$\jmat^\mu-m^2 V_\cl^\mu$
is the conserved current, where 
 $-m^2 V^\mu_\cl$ plays the role of the internal current.
The internal current $-m^2 V^\mu_\cl$
is  sometimes  referred to as a 
 pseudocurrent  \cite{Nieto_RMP1971,Nieto_RMP2010}. 
In our earlier works, we comprehensively explored its surprising
influence on the dynamics of the vector field  using 
extreme yet  particularly illustrative examples \cite{BDPeriodic1,BDCounter}.

Fifth,  theory (\ref{LG}) is gauge non-invariant even when the electric charge 
is conserved.
This means that the {\it action} is not invariant 
with respect to 
\be
A^\cl_\mu\to A^\cl_\mu + \partial_\mu\chi,
\ee
where $\chi$ is an ``arbitrary'' function.
Still, the field equations are invariant under  restricted gauge
transformations, which are defined by  $\Box\chi=0$.

Sixth, the energy is
\begin{subequations}
\begin{align}
&H_\cl=H^\cl_0 + H^\cl_1,\\
\label{HH00}
&H_0^\cl=   -\frac{1}{2}\int d^3r \B{\partial_t A_\cl^\mu\partial_t A^\cl_\mu 
+\bnabla A_\cl^\mu\cdot  \bnabla A^\cl_\mu},\\
&H_1^\cl=\int d^3r \jmat^\mu A^\cl_\mu,
\end{align}
\label{HHHH}%
\end{subequations}
where  $H^\cl_0$ ($H^\cl_1$) will be referred to as the free-field
(interaction) energy and $\bnabla=(\partial/\partial r^i)$.
These expressions follow  from (\ref{LG1}); see  \cite{Greiner} for
details.

Seventh, the Maxwell-type formulation of (\ref{elf})  is 
\begin{align}
\label{qqqq1}
&\bnabla\cdot\BD{E}_\cl=\jmat^0  -\partial_t C_\cl,\\
&\bnabla\times\BD{B}_\cl=\BD{j}_\text{mat} + \bnabla C_\cl +
\partial_t\BD{E}_\cl,
\label{qqqq2}
\end{align}
where $\BD{E}_\cl=-\bnabla A^0_\cl - \partial_t\BD{A}_\cl$,
$\BD{B}_\cl=\bnabla\times\BD{A}_\cl$, and the dynamics of the scalar field 
\be
C_\cl=\partial\cdot A_\cl
\ee
is already 
encoded in these equations and agrees with (\ref{Cdyn}).
The remaining equations, 
\be
\bnabla\times\BD{E}_\cl=-\partial_t\BD{B}_\cl, \ \bnabla\cdot\BD{B}_\cl=0,
\label{qqqq3} 
\ee
are just as in the Maxwell theory. 
This representation of the field equations is commonly used in the literature on the theory under
consideration
(see  e.g. \cite{ModaneseSym2023} and references therein).

Having said all that, we can define  the problem that we would
like to solve. Namely, we are interested in finding the 
dynamics of the vector  field and related quantities 
resulting from the following time dependence of the current
\be
\jmat(t,\BD{r})=(\jmat^0(\BD{r}),\0)\theta(t),
\label{jjjm}
\ee
where $\theta$ is the Heaviside step function and we assume that the vector  
field  vanishes  for $t<0$.


\section{Quantum-to-classical mapping}
\label{Quantum_sec}

The problem defined at the end of the previous section can be addressed
by solving field equations (\ref{boxAj}) with  appropriate initial
conditions. 
However, we adopt a different approach. Specifically, we construct a quantum theory 
based on  Lagrangian density (\ref{LG1}) and determine the wavefunction describing 
the system that is initially in the vacuum state and subsequently evolves under (\ref{jjjm}). 
The expectation values of the field operators and of the two components of the 
Hamiltonian reproduce their classical counterparts. Readers interested solely 
in the classical aspects of the problem may omit 
Secs.~\ref{Quantization_sec}--\ref{General_sec}.

\subsection{Quantization}
\label{Quantization_sec}

We canonically quantize the theory 
\be
{\cal L}= -\frac{1}{2} (\partial_\mu A_\nu)^2 - \jmat^\mu A_\mu,
\ee
which leads to
\begin{align}
\label{canCOMM}
&[A^\mu(0,\BD{x}),\pi^\nu(0,\BD{y})]=
\ii\eta^{\mu\nu}\delta(\BD{x}-\BD{y}),\\
\label{canCOMM_AA}
&[A^\mu(0,\BD{x}),A^\nu(0,\BD{y})]=0,\\
&[\pi^\mu(0,\BD{x}) \pi^\nu(0,\BD{y})]=0,
\label{canPiPi}
\end{align}
where $\pi^\mu=\partial \Lag/\partial(\partial_t A_\mu)=-\partial_t A^\mu$.
These relations are satisfied by \cite{Greiner} 
\begin{align}
\label{AIGB}
&A^\mu(0,\BD{r})= \int
\frac{d^3k}{(2\pi)^{3/2}}
\frac{1}{\sqrt{2\OM{k}}}
\sum_\sigma
\eta^\mu(\BD{k},\sigma)a_{\BD{k}\sigma}\exp(\ii\BD{k}\cdot\BD{r})
+\text{h.c.},\\
\label{PIGB}
&\pi^\mu(0,\BD{r})= \ii\int
\frac{d^3k}{(2\pi)^{3/2}}
\sqrt{\frac{\OM{k}}{2}}
\sum_\sigma
\eta^\mu(\BD{k},\sigma)a_{\BD{k}\sigma}\exp(\ii\BD{k}\cdot\BD{r})
+\text{h.c.},\\
&
 \lbrack a_{\BD{k}\sigma},a^\dag_{\BD{k}'\sigma'}\rbrack=\chi(\sigma)\delta_{\sigma\sigma'}\delta(\BD{k}-\BD{k}'),
 \ \lbrack a_{\BD{k}\sigma},a_{\BD{k}'\sigma'}\rbrack=0,\\
 &\chi(\sigma=0)=-1, \ \chi(\sigma=1,2,3)=1,\\
 & \sigma,\sigma'=1,2: \ \eta(\BD{k},\sigma)=(0,\BD{\eta}(\BD{k},\sigma)), \
  \BD{\eta}(\BD{k},\sigma)\in\RR^3, \
\BD{\eta}(\BD{k},\sigma)\cdot\BD{\eta}(\BD{k},\sigma')=\delta_{\sigma\sigma'}, \
\BD{\eta}(\BD{k},\sigma)\cdot\BD{k}=0,\\
& \eta(\BD{k},0)=(1,\0), \ \eta(\BD{k},3)=(0,\hat{\BD{k}}),
\end{align}
where $\OM{k}=|\BD{k}|$, $\sum_\sigma\equiv\sum_{\sigma=0}^3$, and all operators $a_\KS$ annihilate the
vacuum state $|0\ra$.
Moreover, using the above formulas, one finds  the following expression for the
  normal-ordered free-field and interaction 
  Hamiltonians in the Schr\"odinger picture
\begin{align}
\label{H0free}
& H_0 = \int d^3k \OM{k} \sum_\sigma \chi(\sigma)a^\dag_\KS a_\KS,\\
& H_1= \int \frac{d^3k}{(2\pi)^{3/2}}\frac{1}{\sqrt{2\OM{k}}}
\sum_\sigma J^*_\text{mat}(t,\BD{k})\cdot\eta(\BD{k},\sigma)
a_\KS+\hc,
\label{H1inte}
\end{align}
where  $J^\mu_\text{mat}(t,\BD{k})= \int d^3r \jmat^\mu(t,\BD{r})
\exp(-\ii\BD{k}\cdot\BD{r})$; $H=H_0+ H_1$.
Note that Schr\"odinger-picture interaction Hamiltonian
(\ref{H1inte}) is time dependent whenever the external current
varies with time.

\subsection{Commutation relations}

The Lorenz-gauge condition is incompatible with canonical 
commutation relations (\ref{canCOMM})--(\ref{canPiPi}).
This is  seen if one combines 
$\ii\delta(\BD{x}-\BD{y})= [A^0(0,\BD{x}),-\partial_t A^0(0,\BD{y})]$,
which follows  from (\ref{canCOMM}),
with $\partial_t A^0(0,\BD{y})=-\partial_i A^i(0,\BD{y})$. The resulting 
equation  violates (\ref{canCOMM_AA}), which can be also seen as a
violation  of microcausality (commutativity of spacelike-separated 
vector fields in the discussed context). By the same token, 
one may start from $[\pi^j(0,\BD{x}),-\partial_t A^0(0,\BD{y})]=0$
following from 
(\ref{canPiPi}) and arrive  at the result contradicting (\ref{canCOMM}).
A similar discussion can be found in \cite{Greiner}.

\subsection{Photon content}
\label{Photon_sub}

Scalar, transverse, and longitudinal photon modes
correspond to $\sigma=0$, $\sigma=1,2$, and $\sigma=3$, respectively.
In total, there are four photon polarizations, which is 
a maximum number in four-dimensional spacetime.
This feature reflects the fact that there is no gauge condition
imposed on a vector field operator by the
quantization procedure, just as there is no such condition imposed
on the classical vector field in theory (\ref{LG1}).
Further insights into the  photon content of the theory are  provided by 
the electric, $\BD{E}=-\bnabla A^0 - \partial_t \BD{A}$, and magnetic,
$\BD{B}=\bnabla\times\BD{A}$, field operators. Without going into details,
we mention that creation and annihilation operators for all 
four photon polarizations  are present in the electric field operator.
However, in the magnetic field operator one finds such operators
for transverse polarizations  only. 
Thereby,  scalar and longitudinal 
photons, unlike their transverse counterparts,  
carry  an   electric field only.

\subsection{Fock and coherent states}
\label{Fock_sec}

We switch to  the  quantization-in-a-periodic-box 
setup, where $V$ denotes the  volume of space in which the theory is quantized,
and summarize properties of Fock and coherent states
for an arbitrary photon polarization.
The mapping between the  continuous and discrete versions of the theory 
is  realized     via standard expressions
\be
\int \frac{d^3k}{(2\pi)^{3/2}}a_\KS\leftrightarrow \sum_\BD{k} \frac{b_\KS}{\sqrt{V}},
\ \int d^3k a^\dag_\KS a_\KS \leftrightarrow  \sum_\BD{k} b^\dag_\KS b_\KS, \
\int \frac{d^3k}{(2\pi)^3}
\leftrightarrow
\sum_{\BD{k}}\frac{1}{V},
\label{BOX}
\ee
where the  new  operators satisfy $[b_\KS,b^\dag_{\BD{k}'\sigma'}]=\chi(\sigma)\delta_{\sigma\sigma'}  \delta_{\BD{k}\BD{k}'}$
and $[b_\KS,b_{\BD{k}'\sigma'}]=0$. Moreover, $|0\ra=\bigotimes_\KS|0_\KS\ra$
and $b_\KS|0_\KS\ra=0$.\footnote{The tensor product here and in (\ref{SchSch}) is taken over all momenta
$\BD{k}$ and polarizations $\sigma=0,1,2,3$; the same convention applies to sums over $\BD{k}\sigma$.}

Fock states,
\be
|n_\KS\ra= \frac{\B{b_\KS^\dag}^n}{\sqrt{n!}}|0_\KS\ra,
\ee
have the following properties 
\begin{align}
&b_\KS|n_\KS\ra = \chi(\sigma)\sqrt{n}|(n-1)_\KS\ra, \\
\label{nKS}
&\lara{n_\KS|n_\KS}=
\chi^n(\sigma), \\
&n_\KS |n_\KS\ra = n |n_\KS\ra,
\end{align}
where $n_\KS=\chi(\sigma) b^\dag_\KS b_\KS$.
As is seen from (\ref{nKS}), Fock states do not belong to Hilbert space.
They belong to indefinite-metric space, where a  norm of a 
state--$\lara{\psi|\psi}$ for a  state $|\psi\ra$ being a  superposition
of Fock states--can be positive,   negative, or null (see e.g. \cite{Greiner}).

Then, we introduce the shift operator
\be
\label{shiftD}
D(\gamma_\KS)=\exp\B{\gamma_\KS b^\dag_\KS - \gamma^*_\KS b_\KS}
\ee
having the following  properties that can be proved via 
Baker–Campbell–Hausdorff-like relations
\begin{align}
& D(\gamma_\KS)=\exp\B{-\chi(\sigma)
|\gamma_\KS|^2/2}\exp\B{\gamma_\KS b^\dag_\KS}\exp\B{-\gamma^*_\KS b_\KS},\\
\label{DbD}
& D^\dag(\gamma_\KS) b_\KS D(\gamma_\KS) = b_\KS +\chi(\sigma) \gamma_\KS.
\end{align}
Coherent states,
\be
\label{cohDD}
 |\gamma_\KS\ra =  D(\gamma_\KS)|0_\KS\ra,
\ee
 have the following properties
\begin{align}
\label{cohAMP}
&|\gamma_\KS\ra= \exp\B{-\chi(\sigma) |\gamma_\KS|^2/2}
\sum_{n=0}^\infty \frac{\gamma_\KS^n}{\sqrt{n!}}|n_\KS\ra,\\
\label{COHH}
&b_\KS |\gamma_\KS\ra=
\chi(\sigma) \gamma_\KS |\gamma_\KS\ra,\\
\label{cohNORM}
&\lara{\gamma_\KS|\gamma_\KS}=1.
\end{align}
These expressions generalize standard Hilbert-space quantum optics formulas,
which are discussed in \cite{Gerry,Scully}, to indefinite-metric space. 
Coherent states in indefinite-metric space
arise in quantum electrodynamics.
Notably,  in the context of the Kulish–Faddeev 
dressing of bare electron and positron states by long-wavelength photons 
\cite{KulishFaddeev}; see also \cite{FurugoriPRD2021} for a recent discussion.

\subsection{General results}
\label{General_sec}

Broadly speaking, the primary aim of this work is to analyze 
the dynamics induced by (\ref{jjjm}). 
However, with 
little additional effort, we can present  in this section  expressions 
applicable to a more general setting in which 
the spacelike components 
of the current do not necessarily vanish
\be
\jmat(t,\BD{r})=\jmat(\BD{r})\theta(t).
\label{jmattt}
\ee
This generalization offers certain advantages.
First, contributions from all four photon polarizations appear explicitly, 
thereby clarifying the similarities and differences between transverse and 
non-transverse  components of  the quantities under consideration.
Second, owing to their generality,
the resulting expressions may be applied to  other problems as well.

The following results are derived by applying to the present problem the 
procedure discussed in detail in \cite{BDmag}, which is itself a generalization
of the textbook results on coherent-state generation by an external current \cite{Gerry,Scully}.
Since this can be achieved 
rather straightforwardly, we present only the final expressions
without dwelling on their derivation.
To proceed, we assume that the system is initially ($t=0$) in the vacuum state
$|0\ra$ and that   (\ref{jmattt}) holds.
We do not specify  any particular form of 
$\jmat(\BD{r})$ from (\ref{jmattt}). We assume only that it  is such
that the quantities under consideration are finite and that the Riemann-Lebesgue
lemma is applicable  whenever it is invoked.
The Riemann-Lebesgue lemma in question reads
\be
\lim_{t\to\infty}\int_0^\infty d\OM{k} f(\OM{k})\exp(\pm\ii\OM{k}t)=0,
\label{kldsf}
\ee
where $f\in L^1(0,\infty)$ is real-valued.
As we deal with sums over $\BD{k}$ below, one has to first
replace $\sum_{\BD{k}}$ by $\int d^3k$, and 
then perform the angular integration in momentum space 
in order to use (\ref{kldsf}).

The Schr\"odinger-picture wavefunction for $t>0$
reads\footnote{A similar formula also holds when the suddenly 
switched-on 
external current is arbitrarily time dependent; see Appendix \ref{General_app}.}
\begin{align}
\label{SchSch}
&|\psi(t)\ra= \bigotimes_\KS \exp\B{\ii\text{Im}\Phi_\KS(t)} 
|\alpha_\KS(t)\exp(-\ii\OM{k}t) \ra,\\
\label{mdoi1}
&\Phi_\KS(t)=\chi(\sigma)\B{1+\ii \OM{k}t -\exp(\ii
\OM{k}t)}|\lambda_\KS|^2,\\
\label{mdoi2}
&\alpha_\KS(t)=(\exp(\ii\OM{k}t)-1)\lambda_\KS,\\
&\lambda_\KS=-\frac{J_\text{mat}(\BD{k})\cdot\eta(\BD{k},\sigma)}{\sqrt{2V}\OM{k}^{3/2}},\\
&J_\text{mat}(\BD{k})=\int d^3r
\jmat(\BD{r})\exp(-\ii\BD{k}\cdot\BD{r}),
\end{align}
which generalizes to the case of four photon polarizations the expressions
from \cite{BDmag} (two photon polarizations) and \cite{BDProca}
(three vector-boson polarizations); note that the problems studied 
in \cite{BDmag,BDProca} 
are formulated in a Hilbert-space setting unlike the problem discussed  here.
Note also that (\ref{SchSch}) satisfies  $\la\psi(t)|\psi(t)\ra=1$.
Moreover, the vector field,
the free-field Hamiltonian, and the interaction Hamiltonian, all
in the Schr\"odinger picture, are  
given by (\ref{AIGB}), (\ref{H0free}), and (\ref{H1inte}), respectively.
Several quantities can be discussed now.

For the vector field, we find for $t>0$ 
\begin{align}
\label{Avecrt}
&\lara{\psi(t)|A^\mu(0,\BD{r})|\psi(t)}=\tilde{A}^\mu_\cl(\BD{r})
-\sum_\KS
\frac{\chi(\sigma)\eta^\mu(\BD{k},\sigma)}{\sqrt{2V\OM{k}}}
\B{\lambda_\KS\exp(-\ii\OM{k}t+\ii\BD{k}\cdot\BD{r})+\cc},\\
\label{Amucl}
&\tilde{A}^\mu_\cl(\BD{r})=\sum_\KS
\frac{\chi(\sigma)\eta^\mu(\BD{k},\sigma)}{\sqrt{2V\OM{k}}}\lambda_\KS
\exp(\ii\BD{k}\cdot\BD{r})  +\cc,
\end{align}
where 
$\tilde{A}^\mu_\cl(\BD{r})$ represents the classical static vector field in the presence
of a perpetually static current $\jmat^\mu(\BD{r})$
\be
-\Delta \tilde{A}^\mu_\cl(\BD{r})=\jmat^\mu(\BD{r});
\label{Astat}
\ee
verification of (\ref{Astat}) is facilitated by (\ref{compL}).
Moreover, invoking the 
 Riemann-Lebesgue lemma, we have
 \be
\lim_{t\to\infty}\lara{\psi(t)|A^\mu(0,\BD{r})|\psi(t)}=\tilde{A}^\mu_\cl(\BD{r})
\label{ain}
\ee
when the limit is taken  for  fixed $\BD{r}$.

For the free-field and interaction energies, we find  for $t>0$  
\be
\label{H0mH1}
\lara{\psi(t)|H_0|\psi(t)}=2\sum_\KS \chi(\sigma) \OM{k}|\lambda_\KS|^2  \B{1-\cos(\OM{k}t)}
=-\lara{\psi(t)|H_1|\psi(t)},
\ee
implying 
\be
\lara{\psi(t)|H|\psi(t)}=\lara{\psi(t)|H_0|\psi(t)} + \lara{\psi(t)|H_1|\psi(t)}=0. 
\label{HZeroo}
\ee
In the long-time limit, 
we invoke the Riemann-Lebesgue lemma once again arriving at 
\begin{align}
\label{H0inf}
&\lim_{t\to\infty}\lara{\psi(t)|H_0|\psi(t)}=2\sum_\KS \chi(\sigma) \OM{k}|\lambda_\KS|^2
=-2 \tilde{H}_\cl,\\
\label{H1inf}
&\lim_{t\to\infty}\lara{\psi(t)|H_1|\psi(t)}=2 \tilde{H}_\cl,
\end{align}
where   
\begin{align}
\label{H01cl}
&\tilde{H}_\cl= \tilde{H}_0^\cl + \tilde{H}_1^\cl,\\
\label{H0cl}
&\tilde{H}_0^\cl = -\frac{1}{2} \int d^3r 
\bnabla \tilde{A}_\cl^\mu(\BD{r})\cdot \bnabla \tilde{A}^\cl_\mu(\BD{r})
=-\tilde{H}_\cl,\\
\label{H1cl}
&\tilde{H}_1^\cl = \int d^3r \jmat^\mu(\BD{r}) \tilde{A}^\cl_\mu(\BD{r})=2 \tilde{H}_\cl 
\end{align}
with  the tilde  again  denoting  static quantities. 
Results (\ref{H0cl}) and (\ref{H1cl}) trivially follow from (\ref{Astat}).
They have the virial theorem-like  structure. As a similar 
property  was 
observed in \cite{BDmag,BDProca}, we shall not dwell on it.

\subsection{Classical results}
\label{Cl_sec}
The quantum-to-classical mapping is 
\be
A^\mu_\cl(t,\BD{r})= \lara{\psi(t)|A^\mu(0,\BD{r})|\psi(t)},
\label{AAaa}
\ee
where $|\psi(t)\ra$ is a Schr\"odinger-picture wavefunction.
To see that it is the case, 
one may compute the expectation value of the
Heisenberg-picture 
field equations, $\Box A^\mu(t,\BD{r})=\jmat^\mu(t,\BD{r})$, in
the Heisenberg-picture state $|\psi(0)\ra$. Then, one 
may  note that $\lara{\psi(0)|A^\mu(t,\BD{r})|\psi(0)}$
is given by (\ref{AAaa}), and so $A^\mu_\cl(t,\BD{r})$ 
 satisfies 
(\ref{boxAj}).
Thereby, $A^\mu_\cl(t,\BD{r})$ is given by (\ref{Avecrt}),
and so for $t>0$  it can be written
as\footnote{\label{Cauchy}The extension of this vector 
field to all times is achieved by multiplying it by
$\theta(t)$. The resulting vector field satisfies null Cauchy initial
conditions: the vector field itself  and its 
time derivative vanish at $t=0$ (the latter 
in the distributional sense).
The same remark applies to the solution discussed in \cite{BDmag}, 
where this was not noted.
}
\begin{align}
\label{amucl}
&A^\mu_\cl(t,\BD{r})=\tilde{A}^\mu_\cl(\BD{r})-\int \frac{d^3k}{(2\pi)^3}
\BB{\frac{J^{\mu}_\text{mat}(\BD{k})}{2\OM{k}^2}\exp(-\ii\OM{k}t+\ii\BD{k}\cdot\BD{r})+\cc},\\
&\tilde{A}^\mu_\cl(\BD{r})=\int \frac{d^3k}{(2\pi)^3}
\frac{J^{\mu}_\text{mat}(\BD{k})}{2\OM{k}^2}\exp(\ii\BD{k}\cdot\BD{r})+\cc,
\label{AsT}
\end{align}
where the sums over $\BD{k}$ have been  replaced by the integrals
 and
the completeness relation in Minkowski space,
\be
\sum_\sigma \chi(\sigma)\eta^\mu(\BD{k},\sigma)
\eta^\nu(\BD{k},\sigma)=-\eta^{\mu\nu},
\label{compL}
\ee
has been  employed
to carry out summations over photon polarizations.
For the sake of completeness,
we mention that  static vector field  (\ref{AsT}) satisfies (\ref{Astat})
and that
$\lim_{t\to\infty}A^\mu_\cl(t,\BD{r})=\tilde{A}^\mu_\cl(\BD{r})$ for fixed
$\BD{r}$ due to (\ref{ain}).

Then, we note that for $t>0$ 
\begin{align}
\label{H0class}
&H_0^\cl(t)   = \lara{\psi(t)|H_0|\psi(t)},\\
&H_1^\cl(t)   =\lara{\psi(t)|H_1|\psi(t)}.
\label{H1class}
\end{align}
The latter result is obvious while the former is explained 
 in Appendix \ref{Derivation_app}.
Proceeding similarly as above, we find  that  
(\ref{H0mH1})--(\ref{H1inf}) 
amount to 
\begin{align}
\label{h0cl}
&H_0^\cl(t)= -\int
\frac{d^3k}{(2\pi)^3}\frac{1-\cos(\OM{k}t)}{\OM{k}^2}J_\text{mat}(\BD{k})\cdot J^*_\text{mat}(\BD{k})
=-H_1^\cl(t),\\
&H_\cl=H_0^\cl(t)+H_1^\cl(t)=0,\\
\label{H0clinf}
&\lim_{t\to\infty}H_0^\cl(t)=-2\tilde{H}_\cl,\\
\label{H1clinf}
&\lim_{t\to\infty}H_1^\cl(t)=2\tilde{H}_\cl,\\
&\tilde{H}_\cl= \int
\frac{d^3k}{(2\pi)^3}\frac{J_\text{mat}(\BD{k})\cdot J^*_\text{mat}(\BD{k})}{2\OM{k}^2},
\label{HCL}
\end{align}
which are valid for $t>0$.
Several remarks are in order now.

First, we take a close look at the infrared  structure of the expanding
 and static fields. Taking into account that
 $J^*_\text{mat}(\BD{k})=J_\text{mat}(-\BD{k})$, we rewrite (\ref{amucl}) to 
 the following form
 \be
A^\mu_\cl(t, \BD{r})=\int \frac{d^3k}{(2\pi)^3}
\frac{1-\cos(\OM{k}t)}{\OM{k}^2}J^{\mu}_\text{mat}(\BD{k})\exp(\ii\BD{k}\cdot\BD{r}),
\label{lwdwneodi}
\ee
and observe that the expression describing 
the corresponding static field $\tilde{A}^\mu_\cl(\BD{r})$ is given by
(\ref{lwdwneodi}) without the factor $1-\cos(\OM{k}t)$. 
This shows that for $\OM{k}\lesssim 1/t$, the Fourier components of the expanding
field are suppressed by the factor $(\OM{k}t)^2/2$ relative 
to those of the static field.
We interpret this suppression as a consequence 
of the causal disappearance of the expanding vector
field for $r\gtrsim t$.
Such an  observation, which was not recognized in \cite{BDmag},
highlights the emergence of the dynamical length
scale $t$ in the expanding field.

Second, we comment on the disappearance in the above expressions of the factor
$\chi(\sigma)$ that appears in all expectation values discussed in 
Sec. \ref{General_sec}.
This factor, originating from the indefinite metric of the state space,
is absorbed by  completeness relation of Minkowski spacetime (\ref{compL}), 
which is itself characterized by an indefinite metric. This 
observation  illustrates a striking interplay between
the well-established indefinite-metric character of Minkowski spacetime and 
the  problematic  indefinite-metric  character   of the state space.

Third,  the energy vanishes for both $t>0$ and  $t<0$ (the 
external current is switched-on ``too fast'' to modify the system's  energy).
This result does not imply that the studied field configuration can be 
spontaneously created because it 
does not account for the energy required to create the  field source.

Fourth, we observe that   (\ref{H0clinf}) and (\ref{H0cl}) disagree while 
(\ref{H1clinf}) and (\ref{H1cl}) agree. This has a simple explanation if 
one assumes that there is a shockwave-like front generated  by the
appearance of the external current and   propagating at the speed of light.
Indeed, in this case (\ref{H1clinf}) is insensitive to its presence,
which explains the agreement with (\ref{H1cl}),
because for a spatially localized $\jmat(\BD{r})$ the effective supports
of the shockwave-like front and the current  do not overlap in the long-time limit.
Thereby, the energy associated with the shockwave-like front must 
be  deposited in  (\ref{H0clinf}), which
accounts for the  disagreement with (\ref{H0cl}). 
This  explanation is consistent with both the following
discussion and our earlier work \cite{BDmag}.

Finally, for the sake of completeness, we provide in Appendix
\ref{General_app} the solution describing  the vector field
after the sudden switch-on of  an arbitrarily
time-dependent external current.


\section{Expanding Coulomb field}
\label{Expanding_sec}

We now return to the problem formulated at the end of Sec. \ref{Classical_sec}
and focus on the dynamics induced by
\begin{align}
\label{jmatext}
&\jmat(t,\BD{r})=\jmat(\BD{r})\theta(t),\\
\label{jmatr}
&\jmat(\BD{r})= \B{q\delta_\epsilon(\BD{r}),\0}, \\ 
&\delta_\epsilon(\BD{r})=
\frac{\exp\B{-r^2/\epsilon^2}}{(\sqrt{\pi}\epsilon)^3},
\label{delta3D}
\end{align}
where $\delta_\epsilon(\BD{r})$ is the three-dimensional  nascent delta function, $r=|\BD{r}|$, 
and $\epsilon>0$
is the characteristic (linear) size of the newborn charge.
This leads to 
\begin{align}
&J_\text{mat}(t,\BD{k})=J_\text{mat}(\BD{k}) \theta(t),\\
&J_\text{mat}(\BD{k})=\B{q\exp\B{-\frac{\epsilon^2\OM{k}^2}{4}},\0}.
\label{JJMAT}
\end{align}
Vanishing of the spacelike components of 
(\ref{JJMAT}) implies  $ \BD{A}_\cl(t,\BD{r})=\0$,
and so  $\BD{B}_\cl(t,\BD{r}) =\0$.

We  work below with  the error function
\be
\erf(x)=\frac{2}{\sqrt{\pi}}\int_0^x dy \exp(-y^2),
\ee
having the following useful properties 
\begin{align}
&\frac{d}{dx}\erf(x) = \frac{2}{\sqrt{\pi}}\exp(-x^2),\\
&\erf(x) =\sign(x) \for x\to\pm\infty,
\label{erf0}
\end{align}
where $\sign$ is the sign function returning $0,\pm1$.
We provide below several analytical results characterizing
static and expanding problems.
In the course of their derivation, the following 
integrals lead to the final results
\begin{align}
&\int
d\Omega(\khat)\exp(\ii\BD{k}\cdot\BD{r})=4\pi\frac{\sin(\OM{k}r)}{\OM{k}r},\\
\label{uuuu}
& \int_0^\infty dx \frac{\sin(x\alpha)}{x} \exp\B{-\frac{x^2}{4}}
=\frac{\pi}{2} \erf(\alpha),\\
& \int_0^\infty dx  \cos(x\alpha) \exp\B{-\frac{x^2}{2}}=
\sqrt{\frac{\pi}{2}}\exp\B{-\frac{\alpha^2}{2}}.
\label{dddd}
\end{align}
They are obtained  from formulas 3.915.1, 3.952.6, and 3.896.4 of  \cite{Ryzhik};
$\Omega(\khat)$ denotes  a solid angle in momentum 
space. We mention  in passing that (\ref{dddd}) follows from (\ref{uuuu}),
which is evident upon differentiating (\ref{uuuu})  with respect to $\alpha$
and performing straightforward manipulations.

\subsection{Static field}
\label{Static_sec}

The static Coulomb field, in the presence of 
external current (\ref{jmatr}),
is  also of interest.
Using (\ref{AsT}) and (\ref{JJMAT}), we arrive at 
\be
\tilde{A}^0_\cl(\BD{r})=\frac{q}{4\pi r}\erf\B{\frac{r}{\epsilon}}
\label{a0cl}
\ee
and $\tilde{\BD{A}}_\cl(\BD{r})=\0$. This leads to  $\tilde{C}_\cl(\BD{r})=0$ and  
\be
\tilde{\BD{E}}_\cl(\BD{r})=
\frac{q\rhat}{4\pi r^2}\erf\B{\frac{r}{\epsilon}}
- \frac{q\hat{\BD{r}}}{2\pi r}\delta_\epsilon(r),
\ee
where 
\be
\delta_\epsilon(x)=\frac{\exp\B{-x^2/\epsilon^2}}{\sqrt{\pi}\epsilon}
\ee
is the one-dimensional nascent delta function that should not be confused with 
its three-dimensional counterpart (\ref{delta3D}). Note that we keep using 
the tilde to label static quantities.

For $r\to0^+$ at fixed  $\epsilon$, we get 
\be
\tilde{A}_\cl^0(\BD{r}) \simeq \frac{q}{2\pi^{3/2}\epsilon},
\ \tilde{\BD{E}}_\cl(\BD{r})\simeq  \frac{q \BD{r}}{3\pi^{3/2}\epsilon^3},
\label{Stat0}
\ee
which shows that the Coulomb singularity is taken care of by working
with the extended-charge model.
For $r\gg\epsilon$, $\erf(r/\epsilon)\simeq1$ and $\delta_\epsilon(r)\simeq0$,
and so the standard  Coulomb field results are recovered.

Then,  we have
\be
\tilde{H}_\cl= \frac{q^2}{2(2\pi)^{3/2}\epsilon},
\label{HHcl}
\ee
which has been obtained via (\ref{HCL}) and (\ref{JJMAT}).
The same static result can be also obtained 
from a more familiar expression that is explicitly positive:
$\int d^3r  \tilde{\BD{E}}_\cl(\BD{r})\cdot\tilde{\BD{E}}_\cl(\BD{r})/2$,
where  
the contribution coming from  the vector field--external current coupling
is no longer manifest. 

Finally, we emphasize that the Coulomb- and Lorenz-gauge Maxwell
theories yield the same static results as those reported above.

\begin{figure}[t]
\includegraphics[width=\pref\columnwidth,clip=true]{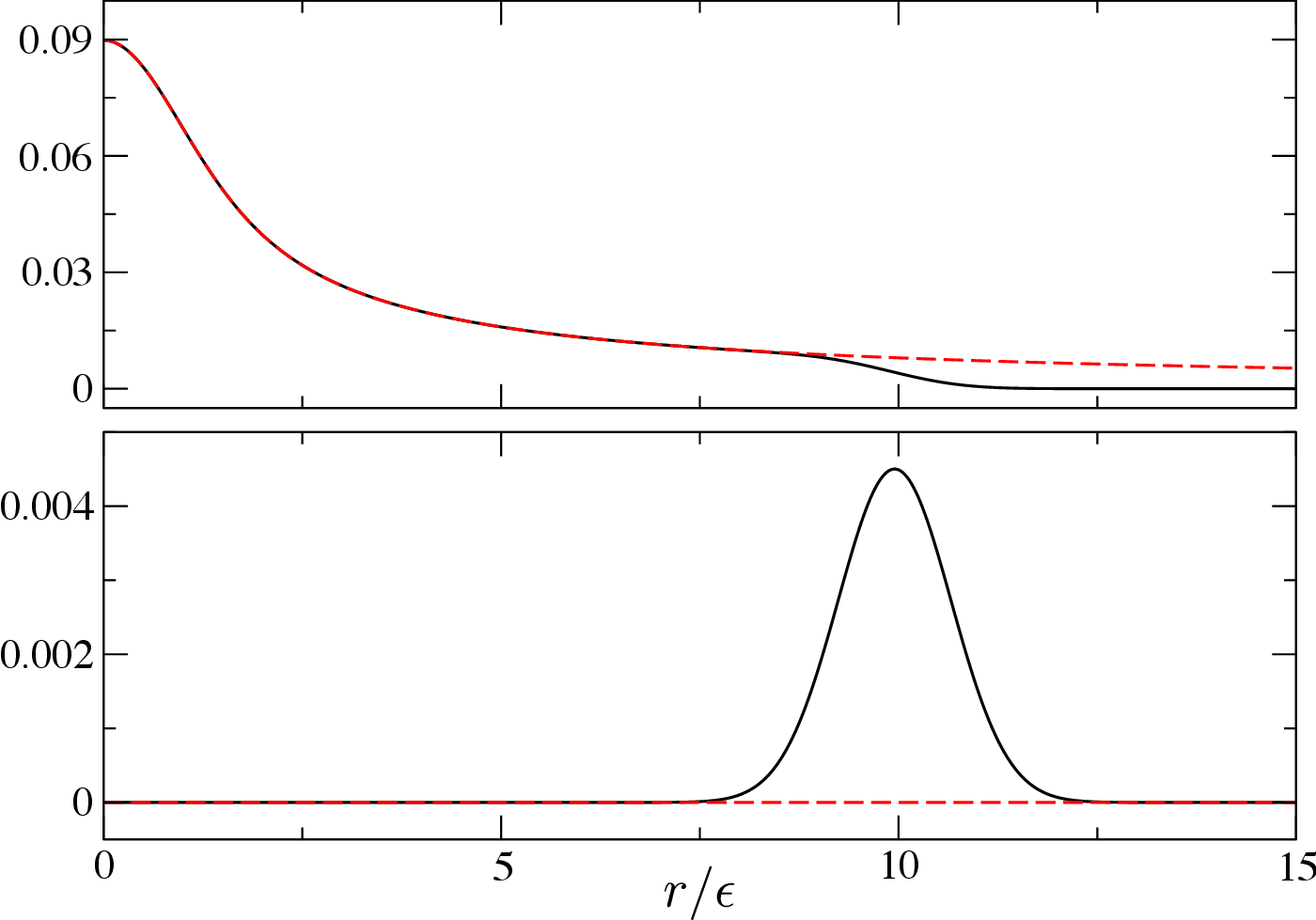}
\caption{Upper panel: the black solid line shows  $\epsilon\times A_\cl^0(t,\BD{r})/q$
for $t/\epsilon=10$, whereas the red dashed line depicts  $\epsilon\times \tilde{A}_\cl^0(\BD{r})/q$.
These lines are practically  indistinguishable on the plotted scale
for $r/\epsilon < t/\epsilon-2$.
Lower panel:  the black solid line shows $\epsilon^2\times C_\cl(t,\BD{r})/q$
for $t/\epsilon=10$ while 
the red dashed line provides the vanishing  static result.
These lines closely overlap for $|r/\epsilon-t/\epsilon|>5/2$.
}
\label{front_fig}
\end{figure}

\subsection{Expanding field}
\label{ExpFil_sub}

For $t>0$, we obtain from (\ref{amucl})  and (\ref{JJMAT})
\be
\label{ext1}
 A_\cl^0(t,\BD{r})  = \frac{q}{4\pi r}
\BB{\erf\B{\frac{r}{\epsilon}}
-\frac{1}{2}\erf\B{\frac{r-t}{\epsilon}}
-\frac{1}{2}\erf\B{\frac{r+t}{\epsilon}}}
\ee
leading to 
\be
 \BD{E}_\cl(t,\BD{r})  =
\frac{q\rhat}{4\pi r^2}\BB{\erf\B{\frac{r}{\epsilon}}
-\frac{1}{2}\erf\B{\frac{r-t}{\epsilon}}
-\frac{1}{2}\erf\B{\frac{r+t}{\epsilon}}}
+ \frac{q\hat{\BD{r}}}{4\pi r}\BB{\delta_\epsilon(r-t)-2\delta_\epsilon(r)+\delta_\epsilon(r+t)}.
\label{ext3}
\ee
These results can be discussed  in the following limiting cases.

For  $t\to0^+$, $A_\cl^0(t,\BD{r})$ and $\BD{E}_\cl(t,\BD{r})$
vanish.
For $r\to0^+$ at fixed $t$ and $\epsilon$,  
\begin{align}
&A_\cl^0(t,\BD{r})\simeq\frac{q}{2\pi}\BB{\delta_\epsilon(0)
-\delta_\epsilon(t)},\\
& \BD{E}_\cl(t,\BD{r})\simeq \frac{q \BD{r}}{3\pi\epsilon^2}
\BB{\delta_\epsilon(0)-\delta_\epsilon(t)+2\delta_\epsilon(t)
\B{\frac{t}{\epsilon}}^2},
\end{align}
reproducing  static (\ref{Stat0}) for $t\gg\epsilon$.
Finally, for $r\gg\epsilon$, 
\begin{align}
&  A_\cl^0(t,\BD{r})  \simeq \frac{q}{8\pi r}
\BB{1-\erf\B{\frac{r-t}{\epsilon}}},\\
& \BD{E}_\cl(t,\BD{r})  \simeq
 \frac{q\rhat}{8\pi r^2}\BB{1-\erf\B{\frac{r-t}{\epsilon}}} 
 + \frac{q\hat{\BD{r}}}{4\pi r}\delta_\epsilon(r-t),
\end{align}
where the expanding nature of the discussed solution is evident. 
Indeed, there is a shockwave-like  front
localized in the {\it neighborhood} of    $r=t$,
which separates the field-free region ($r\gtrsim t+\epsilon$) from 
the Coulomb-field region ($r \lesssim t-\epsilon$).\footnote{We use the 
term shockwave-like because in the limit of $\epsilon\to0^+$,
there is a singular structure in the vector field
that preserves its shape during propagation
(there is no self-steepening dynamics as in actual shockwaves).}
In the former, vanishing of the fields reflects the causal nature 
the charge-creation problem.
In the latter, the static results of Sec.~\ref{Static_sec} hold.
This is illustrated in Fig. \ref{front_fig}.\footnote{Numerical data for
all figures in this work have been generated using \cite{Mathematica143}.}

\subsection{Modified  Maxwell's equations}
Equations (\ref{qqqq1}) and (\ref{qqqq2}) differ from differential 
Maxwell's
equations when spatial and time derivatives of the scalar field
$C_\cl$ are non-negligible. In the discussed problem,
it follows from (\ref{ext1}) that for $t>0$ 
\begin{align}
C_\cl(t,\BD{r}) &=\frac{q}{4\pi r}
\BB{\delta_\epsilon(r-t)-\delta_\epsilon(r+t)},
\label{ccl}
\end{align}
which for $r\gg\epsilon$ reduces to  
\be
C_\cl(t,\BD{r})  \simeq\frac{q}{4\pi r}\delta_\epsilon(r-t).
\label{nl2ddlk}
\ee
This shows that (\ref{qqqq1}) and (\ref{qqqq2}) 
differ from the differential Maxwell's equations only near  the shockwave-like front.
No matter how small this region is, these differences
profoundly affect long-distance behavior of the vector  field.

\subsection{Modified Gauss' law}
 
Modified Gauss' law (\ref{qqqq1}) holds   in the studied theory.
As in the discussed solution  $\int d^3r \bnabla\cdot\BD{E}_\cl=0$ and 
$\int d^3r \jmat^0=q$, we must have 
$-\int d^3r\partial_t C_\cl=-q$,
which  can be directly   verified  using  (\ref{ccl}). 
Hence, as far as $\int d^3r \bnabla\cdot\BD{E}_\cl$ is
concerned, the charge $-q$  associated with the internal current
cancels the electric charge $q$ encoded in the  external
current. 
In other words, the scalar field  perfectly {\it screens} the electric charge.
This  feature quite generally   follows 
from  modified Gauss' law and causality considerations, it is 
not merely a property of the particular solution studied in this section.
Note that the electric charge--ultimately carried by a
matter field--has a different nature from the internal charge that 
is carried by the vector field. These  two quantities should not be confused.

Finally, as the above-mentioned screening disappears for a time-independent
scalar field, it is an intrinsically dynamical phenomenon.
This is in contrast to Debye screening of  electric charges in plasmas and electrolytes, 
which persists in  ``static'' settings.

\subsection{Energy}
\label{Energy_sub}

The free-field energy of our field configuration,
which is obtained from (\ref{h0cl}) and (\ref{JJMAT}),
reads 
\be
H_0^\cl(t)=
-\frac{q^2}{(2\pi)^{3/2}\epsilon}+
\frac{q^2}{(2\pi)^{3/2}\epsilon}\exp\B{-\frac{t^2}{2\epsilon^2}}
\label{freeH0}
\ee
while the interaction energy is 
$H^\cl_1(t)=-H_0^\cl(t)$; note expected 
vanishing of these quantities for  $t\to0^+$ and the fact that  $t>0$ is
assumed above.
Equivalently, using (\ref{HHcl}), the above result can be written as 
\be
H_0^\cl(t)=-2 \tilde{H}_\cl\BB{1-\exp\B{-\frac{t^2}{2\epsilon^2}}}
\ee
and compared to  (\ref{H0clinf}).
The vanishing of the energy for $t>0$, 
\be
H_0^\cl(t)+H^\cl_1(t)=0,
\label{hzer}
\ee
leads to intriguing insights.

First,  in the spirit of the classical theory of charged
particles \cite{RohrlichClassicalBook}, 
one may use (\ref{hzer}) to argue as follows.
Suppose that the rest energy of a charged particle consists of both
electromagnetic and non-electromagnetic contributions. Then, (\ref{hzer})  
suggests that the creation of  such a particle requires supplying
only the energy associated with the latter contribution.

\begin{figure}[t]
\includegraphics[width=\pref\columnwidth,clip=true]{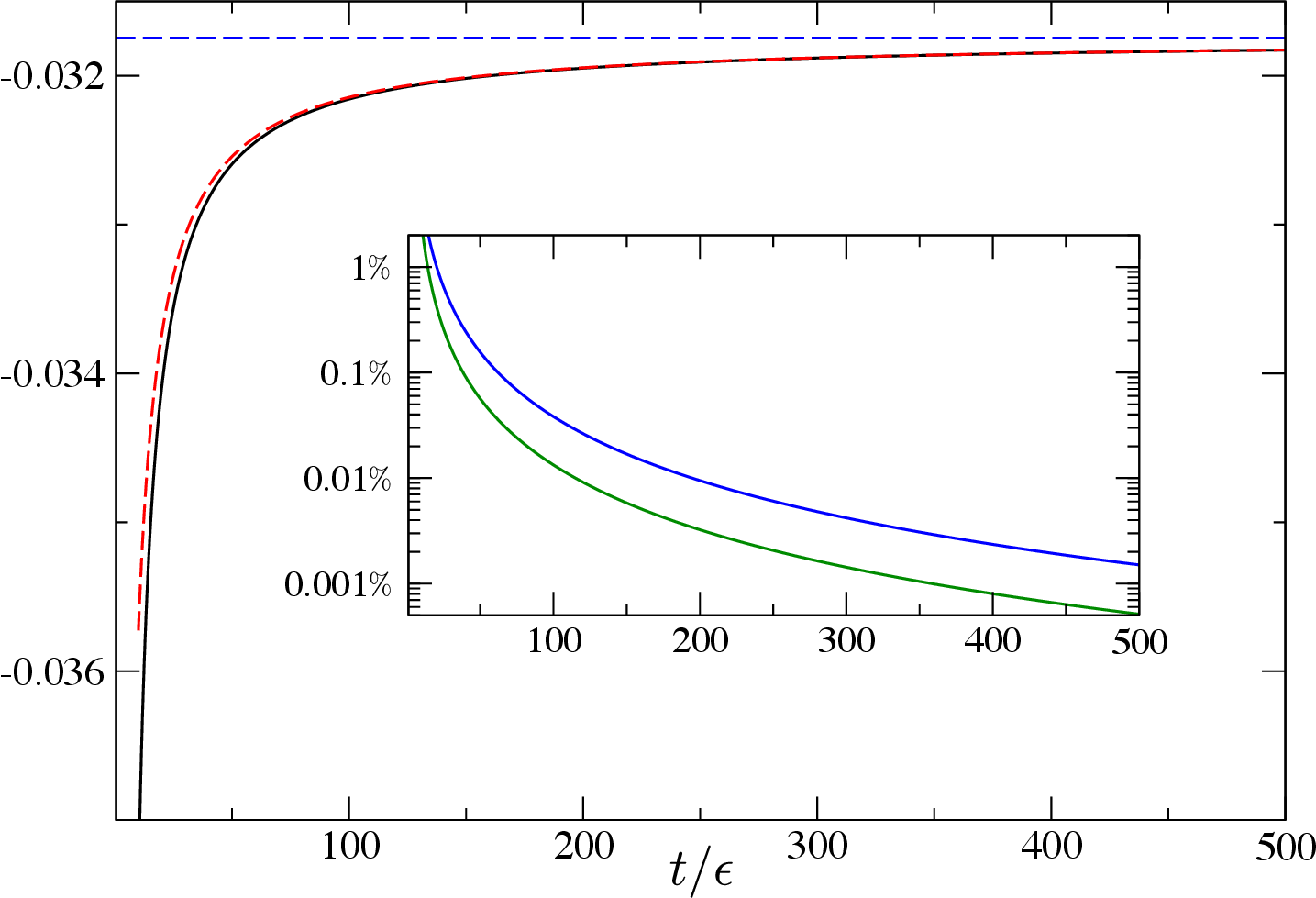}
\caption{{\bf Main plot}. The  black solid line  shows 
$\epsilon\times\int_{\cal V}d^3r{\cal H}_\cl/q^2$, where 
the domain of integration is ${\cal V}=\{\BD{r}:|r-t|<3\epsilon\}$ and 
the energy density ${\cal H}_\cl$ is given by  (\ref{wefkwee}). 
The red and blue dashed lines  show (\ref{lwefowoeew}) and  (\ref{wekdqqq})
multiplied by $\epsilon/q^2$.
The black solid and red
dashed lines are practically  indistinguishable on the plotted scale 
for $t/\epsilon>100$.
{\bf Inset}. 
The blue solid line  shows the  relative difference between 
the black solid and red dashed lines from the main 
plot. The green solid line shows the relative difference between 
the black solid line from the main plot 
and the corresponding data computed with ${\cal V}=\{\BD{r}:|r-t|<4\epsilon\}$.
By the relative difference between 
two quantities, say  $a$ and $b$, we mean $100\%\times |a-b|/|a|$.
}
\label{XX}
\end{figure}

Second, for $t\to\infty$, we have (i) the shockwave-like front  localized
in the region, where the Coulomb field is negligible; (ii) 
the Coulomb field formed in nearly all space.
Remark (i) implies that the energy in this limit is given
by the sum of the Coulomb-field contribution and the shockwave-like front 
contribution. As the former is given by the energy of the static 
Coulomb field (\ref{HHcl}), which follows from remark  (ii), the latter 
must be given by 
\be
-\frac{q^2}{2(2\pi)^{3/2}\epsilon}
\label{wekdqqq}
\ee
to satisfy  (\ref{hzer}).
This  observation can be verified numerically by integrating 
the energy density around $r=t$. 
The results are shown in Fig. \ref{XX}, where besides the numerical data,
we also show the following analytical prediction for the quantity under
consideration 
\be
-\frac{q^2}{2(2\pi)^{3/2}\epsilon} -\frac{q^2}{8\pi t},
\label{lwefowoeew}
\ee
which takes into account a finite-time correction to (\ref{wekdqqq}); see Appendix
\ref{EShock_app} for details.

Third, the above discussion can be generalized to other cases, where $\jmat^0(\BD{r})$ 
is localized but not necessarily given by  $q\delta_\epsilon(\BD{r})$. 
Namely, in the long-time limit, it is expected  that the energy
localized at  the shockwave-like front is 
negative.\footnote{Note that the energy of the Coulomb field of
 a static charge distribution is positive 
(see the discussion at  the end of Sec. \ref{Static_sec}).
Also note that (\ref{hzer}) is not merely a property of the particular solution studied 
in this section (as is discussed  in Sec. \ref{Cl_sec}, this is a quite general result).
}
Since  a shockwave-like front can be seen as  a radiation burst,
such a result is  incompatible with the Maxwell theory.
We see this observation  as an important energy-based insight into
the impossibility of charge creation in the Maxwell theory.

\subsection{Comparison to solution from \cite{AlickiJPA1978}}

The expanding Coulomb field following sudden creation of a point charge
is briefly discussed in \cite{AlickiJPA1978}.
As far as the fields are  concerned,
one may easily argue  that  for $\epsilon\to0^+$, our solution 
is essentially equivalent to the result obtained in \cite{AlickiJPA1978}.
As far as the energy is concerned, the result reported in \cite{AlickiJPA1978} 
is of limited utility because  it 
is mathematically undefined and does not (even formally)
indicate vanishing of the energy. Nonetheless, 
it indicates negativity of
the energy localized at  the shockwave-like front,
which agrees with our findings.
We mention in passing that the solution given in \cite{AlickiJPA1978} was obtained 
directly from a differential equation, which is quite different from our 
``quantum mechanical'' approach.

\subsection{Comparison to expanding magnetic dipole field}
\label{Comparison_sub}

We find it instructive to compare the expanding Coulomb field studied in this work with 
the expanding magnetic dipole field discussed in \cite{BDmag}. 
The latter arises as a response to the external current
$(0,\BD{\mu}\times\BD{r})f(\BD{r})\theta(t)$, 
where $\BD{\mu}$ is a magnetic  moment and $f(\BD{r})=2\delta_\epsilon(\BD{r})/\epsilon^2$.
At first sight, the two problems appear  unrelated because the spacelike 
(timelike) component of the vector field vanishes in the expanding Coulomb 
(magnetic dipole) field. However, the spacelike component of the 
above-listed current can  be also obtained
by acting with the operator $-(\BD{\mu}\times\bnabla)/q$ on the timelike
component of current (\ref{jmatext}).
Thereby, using the field equations, it is easy to argue that the spacelike component of 
the Coulomb-gauge vector field, describing the expanding magnetic dipole
field, must be given by the action of such an operator  on (\ref{ext1}).
This is indeed the case, which   directly confirms the correctness of  
the indefinite-metric-based quantum-to-classical mapping technique
discussed in Sec. \ref{Cl_sec}.

The  mapping between the timelike and spacelike components of the
vector fields, which also applies to static field configurations,
does not imply that the two problems are physically equivalent.
In fact, they are not  because  
{\it different}
theories are employed in their analysis 
(the Maxwell theory used in \cite{BDmag} is incapable of addressing the problem
discussed in this work). This distinction is most clearly seen when one
considers the difference between the energies of the expanding and static field 
configurations (the energy radiated due to the sudden appearance of the current).
It is negative (positive) in the expanding Coulomb  (magnetic dipole) setting. 
Further differences between the two problems are listed 
in Table \ref{tab}.

\begin{table}[t]
\begin{tabular}{c|c}
expanding Coulomb field  & expanding magnetic dipole field \\
\hline
$\partial\cdot\jmat\neq0$ & $\partial\cdot\jmat=0$\\
\hline
expanding electric field & expanding magnetic field\\
\hline
shockwave-localized scalar field & shockwave-localized electric field
\end{tabular}
\caption{Comparison of the   problems discussed in Sec.~\ref{Comparison_sub}.}
\label{tab}
\end{table}


\section{Discussion}
\label{Discussion_sec}

We have discussed a field configuration resulting 
from  electric charge creation. For this purpose, we have 
used  classical  theory (\ref{LG}), 
whose Lagrangian density differs from that of the Maxwell theory by a
term $\propto C_\cl^2$, where  $C_\cl=\partial\cdot A_\cl$. 
This modification of the Lagrangian density 
enables the process of electric charge creation, which is impossible in the Maxwell theory.
We have shown that this process produces a shockwave-like front propagating at the speed of light. 
In the region swept by this front, the  Coulomb field is established, 
whereas in the region not yet reached by it, the Coulomb field is absent due to causality
($C_\cl$ is negligible in both regions). This intuitive picture breaks down  inside 
the shockwave-like front, where $C_\cl\neq0$ and so new physics, absent in the  Maxwell theory, appears. 
In particular, the Coulomb field ``bends'' there from its normal value to
zero, which  is accompanied by a negative energy stored in the  shockwave-like  front.
The question then arises whether such effects can be 
found in Nature. 
In our opinion, the very natural causal picture of the emergence of the Coulomb
field  motivates further investigation of the subject addressed in this work
(e.g.  the impact of the shockwave-like front on charged particles
and its potentially observable consequences).

We have obtained  analytical  solutions of classical field equations 
in a rather peculiar manner. Specifically, we have quantized the
classical theory, solved the time-dependent  Schr\"odinger equation, and associated the 
expectation value of the vector field operator with the classical vector 
field. This raises the question of whether the discussed quantum theory 
is merely  a fancy technical device for studying  a classical problem 
or a  fundamental description of Nature.
While the discussion of this complex issue is beyond the scope of the current
work, we shall comment on the relation between the theory 
discussed in Secs.~\ref{Quantization_sec}--\ref{General_sec} 
and the Gupta-Bleuler (GB) theory of a covariantly-quantized
massless vector  field. 

As far as operators are concerned, the  theory discussed in 
Secs.~\ref{Quantization_sec}--\ref{Photon_sub}  is 
the same as the GB theory. However, in the GB theory,  
one works in the so-called   physical sector of state space \cite{Gupta}. 
A Schr\"odinger-picture wavefunction  $|\Psi(t)\ra$,
belonging to the  GB physical sector, 
satisfies\footnote{In the GB theory, the Lorenz-gauge condition 
 arises  in this   limited sense.}
\be
\lara{\Psi(t)|C(0,\BD{r})|\Psi(t)}=0.
\label{ewmd}
\ee
This property  ensures that 
$\lara{\Psi(t)|\BD{E}(0,\BD{r})|\Psi(t)}$ and 
 $\lara{\Psi(t)|\BD{B}(0,\BD{r})|\Psi(t)}$, which 
can be regarded as a classical 
electromagnetic field, satisfy  the 
Maxwell's equations.
We observe now that
states introduced in Sec.~\ref{Fock_sec} 
have been used in Sec.~\ref{General_sec}
to describe  quantum dynamics. Whenever an electric charge
is non-conserved,
this dynamics takes place  {\it outside} the physical sector of the GB
theory. In fact, the quantum counterpart of the classical dynamics 
discussed  in Sec. \ref{ExpFil_sub} provides a relevant  example of this 
feature.\footnote{This conclusion,  
via the quantum-to-classical mapping, follows 
from the fact that  $C_\cl\neq0$ in the solution presented  in Sec. \ref{ExpFil_sub}.}
We also observe that  theory (\ref{LG}) does not have to possess
an acceptable quantum counterpart in order to be physically useful.
In the absence of such a counterpart, it could still be regarded as
an effective theory that is valid within some regime.
Indeed, there are successful effective classical theories for which no 
{\it direct}   quantum counterpart exists
 (see e.g. classical hydrodynamics).

Finally, we stress that our work assumes  existence of a physical
mechanism leading to electric charge creation. Its description is beyond
the scope of our work, which aims at quantifying the electromagnetic 
fingerprint    of    a   charge-creation process. 
Since this process has not yet been experimentally observed, 
it may be worth considering what its present consequences
would be if it had occurred in the early stages of the Universe's evolution.

\acknowledgments

The research for this publication has been supported by Jagiellonian University.

\appendix


\section{Difficulty encountered in \cite{BermanAmJPhys2008}}
\label{Difficulties_app}

 One of the goals of  \cite{BermanAmJPhys2008} is to discuss  how a 
 Coulomb field could  be dynamically created.
We believe that the proposed 
scenario is physically unacceptable.
Namely, it is assumed in \cite{BermanAmJPhys2008} that 
an electric charge density is given by $q(t)\delta(\BD{r})$, where
$q(t<0)=0$ and  $q(t>0)\neq0$, and that 
the  continuity equation  holds. It is then proposed 
that the 
charge current satisfying this relation is given by $-\dot{q}(t)\BD{r}/(4\pi r^3)$,
where a dot over a variable denotes its time derivative.
This current  suddenly appears in whole space  at $t=0$ (or $t=0^+$
if $\dot{q}(0)=0$), and so it is unphysical (at least on causal grounds).


\section{Conventions}
\label{Conventions_sec}
The Heaviside-Lorentz system of units is used
in its $\hbar=c=1$ version. 
Greek (Latin) tensor indices 
take values $0,1,2,3$ ($1,2,3$). 
The metric signature is $(+---)$.
$3$-vectors are written in bold, e.g. 
$x=(x^\mu)=(x^0,\BD{x})$, and
$\partial_\mu=\partial/\partial x^\mu$.
The Einstein summation convention is used,
$(X_{\mu\cdots})^2=X_{\mu\cdots}X^{\mu\cdots}$.
The complex (hermitian) conjugation is denoted by $*$ and $\cc$ 
($\dag$ and $\hc$). The symbol $\equiv$ denotes a definition, 
$:\,:$ designates the  normal ordering,
$\Delta=\bnabla\cdot\bnabla$ represents the Laplacian,
$\Box=\partial_t^2-\Delta$,  $\partial_t=\partial/\partial t$,
 $x^+$  stands for  a quantity infinitesimally
larger  than $x$,  and $\BD{\hat{x}}=\BD{x}/|\BD{x}|$.


\section{General dynamics}
\label{General_app}

We extend here both the quantum (Sec. \ref{General_sec}) and 
classical (Sec. \ref{Cl_sec})  discussion 
to the case of an arbitrarily time-dependent  external current
that is instantaneously switched on at $t=0$. 

{\bf Quantum dynamics}. 
Assuming that the system is  in 
the vacuum state $|0\ra$ at $t=0$,
the Schr\"odinger-picture wavefunction for $t>0$
is still given by (\ref{SchSch}) but with 
\begin{align}
\label{jiu1}
&\Phi_\KS(t)=\chi(\sigma)\int_0^t dt'\beta_\KS(t')\int_0^{t'} dt''\beta_\KS^*(t''),\\
&\alpha_\KS(t)= \int_0^t dt' \beta_\KS(t'), 
\label{jiu2}
\end{align}
where
\be
\beta_\KS(t)=  \frac{-\ii \exp(\ii\OM{k}t) }{\sqrt{2V\OM{k}}}
J_\text{mat}(t,\BD{k})\cdot\eta(\BD{k},\sigma).
\ee
These results can be obtained by extending the discussion from 
\cite{BDmag} to the indefinite-metric setting.
For  time-independent $J_\text{mat}$,
(\ref{jiu1}) and (\ref{jiu2}) reduce to 
(\ref{mdoi1}) and (\ref{mdoi2}), respectively.

{\bf Classical dynamics}. We introduce 
\be
\widehat{J}_\text{mat}(t,\BD{k})= 
\OM{k}  \int_0^t dt'
J_\text{mat}(t',\BD{k})
\exp[\ii \OM{k}(t'-t)]
\label{hatJ}
\ee
and obtain via  quantum-to-classical mapping (\ref{AAaa}) 
the following 
result for the vector field
\be
A^\mu_\cl(t,\BD{r})=\ii\int\frac{d^3k}{(2\pi)^3}
\frac{ \widehat{J}^{\mu}_\text{mat}(t,\BD{k})     }{2\OM{k}^2}
\exp(\ii\BD{k}\cdot\BD{r})
+\cc,
\label{welkfq}
\ee
which is valid for $t>0$. The remarks formulated in footnote \ref{Cauchy}
apply to this solution as well.
For  time-independent $J_\text{mat}$, (\ref{welkfq})
reduces to (\ref{amucl}).


\section{Derivation of (\ref{H0class})}
\label{Derivation_app}

The Schr\"odinger-picture  counterpart of classical (\ref{HH00}),
expressed in terms of the canonical variables of the theory, reads
\be
H_0=   -\frac{1}{2}\int d^3r 
:\BB{\pi^\mu(0,\BD{r})\pi_\mu(0,\BD{r})+
\bnabla A^\mu(0,\BD{r})\cdot \bnabla A_\mu(0,\BD{r})}:.
\ee
Due to  normal ordering and the fact that the 
normalized-to-unity coherent state 
$|\psi(t)\ra$ is an eigenstate of all annihilation operators,
\begin{align}
\lara{\psi(t)|H_0|\psi(t)}=&-\frac{1}{2}\int d^3r 
\lara{\psi(t)|\pi^\mu(0,\BD{r})|\psi(t)}
\lara{\psi(t)|\pi_\mu(0,\BD{r})|\psi(t)}\nonumber\\
&-\frac{1}{2}\int d^3r
\bnabla\lara{\psi(t)|A^\mu(0,\BD{r})|\psi(t)}\cdot \bnabla 
\lara{\psi(t)|A_\mu(0,\BD{r})|\psi(t)}.
\label{erflk}
\end{align}
Then, we note that (\ref{AAaa}) holds
and   $\lara{\psi(t)|\pi^\mu(0,\BD{r})|\psi(t)}=
\lara{\psi(0)|\pi^\mu(t,\BD{r})|\psi(0)}=
\lara{\psi(0)|-\partial_t A^\mu(t,\BD{r})|\psi(0)}=
-\partial_t A^\mu_\cl(t,\BD{r})$.
With these expressions,  one  sees that (\ref{erflk}) 
is the same as  (\ref{H0class}).


\section{Energy localized near $r=t$}
\label{EShock_app}

In the problem discussed in Secs. \ref{ExpFil_sub}--\ref{Energy_sub},
the energy density  reads 
\be
{\cal H}_\cl=-\frac{1}{2}
\B{C_\cl^2+\BD{E}_\cl\cdot\BD{E}_\cl}+\jmat^0A_\cl^0,
\label{wefkwee}
\ee
which can be argued using  (\ref{HHHH}). For $t\to\infty$ being of interest
here, the last term in this expression 
does not contribute near $r=t$, and so we work below with 
\be
{\cal H}'_\cl=-\frac{1}{2}
\B{C_\cl^2+\BD{E}_\cl\cdot\BD{E}_\cl}.
\ee

Using (\ref{nl2ddlk})  and  
\be
\BD{E}_\cl\simeq  \frac{q\rhat}{4\pi r}\delta_\epsilon(r-t),
\label{eflm12k}
\ee
we  obtain (\ref{wekdqqq}) out of $\int d^3r {\cal H}'_\cl$;
see the remark below (\ref{lne3}) for the origin of (\ref{eflm12k}). Two remarks are in
order now. First, we have not restricted the 
domain of integration
to $|r-t|<O(\epsilon)$ because $\delta_\epsilon(r-t)$ does this  for
us. 
Note that we study here the energy contained in the volume 
occupied by the shockwave-like front, which is localized within the effective support of
$\delta_\epsilon(r-t)$.
Second, to  simplify the calculations,
we have made the replacement
\be
\int_0^\infty dr \to \int_{-\infty}^\infty dr.
\label{m3efk}
\ee

To obtain a better approximation, still at long times, we work with
\be
\label{lne3}
\BD{E}_\cl \simeq \BD{E}_\cl^\text{lead} + \BD{E}_\cl^\text{sub}, \ 
\BD{E}_\cl^\text{lead}=\frac{q\rhat}{4\pi r}\delta_\epsilon(r-t), \ 
  \BD{E}_\cl^\text{sub}= \frac{q\rhat}{8\pi r^2}
\ee
while keeping $C_\cl$ the same as in (\ref{nl2ddlk}); $\BD{E}_\cl^\text{lead}$ 
($\BD{E}_\cl^\text{sub}$) is the leading (subleading) contribution 
 near $r=t$ obtained  from  (\ref{ext3}) in the long-time limit.
Then, we  approximate 
\be
\BD{E}_\cl\cdot\BD{E}_\cl\simeq\BD{E}_\cl^\text{lead}\cdot\BD{E}_\cl^\text{lead}
+2\BD{E}_\cl^\text{lead}\cdot\BD{E}_\cl^\text{sub},
\ee
where  we have omitted   $\BD{E}_\cl^\text{sub}\cdot\BD{E}_\cl^\text{sub}$
because its contribution to the energy localized near $r=t$
is  $\propto1/t^2$, which is  negligible in the studies leading to (\ref{lwefowoeew}).
The integration of the resulting expression for the energy density ${\cal H}'_\cl$ 
is performed just as above and leads to
(\ref{lwefowoeew}). While integrating the term
$\propto\BD{E}_\cl^\text{lead}\cdot\BD{E}_\cl^\text{sub}$,
the factor   $1/r$ under the integral sign  has to be
replaced  by $1/t$.



\begin{thebibliography}{28}%
\makeatletter
\providecommand \@ifxundefined [1]{%
 \@ifx{#1\undefined}
}%
\providecommand \@ifnum [1]{%
 \ifnum #1\expandafter \@firstoftwo
 \else \expandafter \@secondoftwo
 \fi
}%
\providecommand \@ifx [1]{%
 \ifx #1\expandafter \@firstoftwo
 \else \expandafter \@secondoftwo
 \fi
}%
\providecommand \natexlab [1]{#1}%
\providecommand \enquote  [1]{``#1''}%
\providecommand \bibnamefont  [1]{#1}%
\providecommand \bibfnamefont [1]{#1}%
\providecommand \citenamefont [1]{#1}%
\providecommand \href@noop [0]{\@secondoftwo}%
\providecommand \href [0]{\begingroup \@sanitize@url \@href}%
\providecommand \@href[1]{\@@startlink{#1}\@@href}%
\providecommand \@@href[1]{\endgroup#1\@@endlink}%
\providecommand \@sanitize@url [0]{\catcode `\\12\catcode `\$12\catcode
  `\&12\catcode `\#12\catcode `\^12\catcode `\_12\catcode `\%12\relax}%
\providecommand \@@startlink[1]{}%
\providecommand \@@endlink[0]{}%
\providecommand \url  [0]{\begingroup\@sanitize@url \@url }%
\providecommand \@url [1]{\endgroup\@href {#1}{\urlprefix }}%
\providecommand \urlprefix  [0]{URL }%
\providecommand \Eprint [0]{\href }%
\providecommand \doibase [0]{https://doi.org/}%
\providecommand \selectlanguage [0]{\@gobble}%
\providecommand \bibinfo  [0]{\@secondoftwo}%
\providecommand \bibfield  [0]{\@secondoftwo}%
\providecommand \translation [1]{[#1]}%
\providecommand \BibitemOpen [0]{}%
\providecommand \bibitemStop [0]{}%
\providecommand \bibitemNoStop [0]{.\EOS\space}%
\providecommand \EOS [0]{\spacefactor3000\relax}%
\providecommand \BibitemShut  [1]{\csname bibitem#1\endcsname}%
\let\auto@bib@innerbib\@empty
\bibitem [{\citenamefont {Griffiths}(2013)}]{GriffithsBook}%
  \BibitemOpen
  \bibfield  {author} {\bibinfo {author} {\bibfnamefont {D.~J.}\ \bibnamefont
  {Griffiths}},\ }\href@noop {} {\emph {\bibinfo {title} {Introduction to
  Electrodynamics}}},\ \bibinfo {edition} {4th}\ ed.\ (\bibinfo  {publisher}
  {Pearson},\ \bibinfo {year} {2013})\BibitemShut {NoStop}%
\bibitem [{\citenamefont {Maggiore}(2023)}]{MaggioreBook}%
  \BibitemOpen
  \bibfield  {author} {\bibinfo {author} {\bibfnamefont {M.}~\bibnamefont
  {Maggiore}},\ }\href@noop {} {\emph {\bibinfo {title} {A Modern Introduction
  to Classical Electrodynamics}}}\ (\bibinfo  {publisher} {Oxford University
  Press},\ \bibinfo {year} {2023})\BibitemShut {NoStop}%
\bibitem [{\citenamefont {Damski}(2026)}]{BDmag}%
  \BibitemOpen
  \bibfield  {author} {\bibinfo {author} {\bibfnamefont {B.}~\bibnamefont
  {Damski}},\ }\bibfield  {title} {\bibinfo {title} {Expanding quantum magnetic
  field},\ }\href@noop {} {\bibfield  {journal} {\bibinfo  {journal} {Nucl.
  Phys. B}\ }\textbf {\bibinfo {volume} {1032}},\ \bibinfo {pages} {117676}
  (\bibinfo {year} {2026})}\BibitemShut {NoStop}%
\bibitem [{BDP()}]{BDProca}%
  \BibitemOpen
  \href@noop {} {}\bibinfo {note} {B. Damski, Expanding quantum Proca field,
  arXiv:2610.00014 (2026).}\BibitemShut {Stop}%
\bibitem [{\citenamefont {Goldhaber}\ and\ \citenamefont
  {Nieto}(2010)}]{Nieto_RMP2010}%
  \BibitemOpen
  \bibfield  {author} {\bibinfo {author} {\bibfnamefont {A.~S.}\ \bibnamefont
  {Goldhaber}}\ and\ \bibinfo {author} {\bibfnamefont {M.~M.}\ \bibnamefont
  {Nieto}},\ }\bibfield  {title} {\bibinfo {title} {Photon and graviton mass
  limits},\ }\href@noop {} {\bibfield  {journal} {\bibinfo  {journal} {Rev.
  Mod. Phys.}\ }\textbf {\bibinfo {volume} {82}},\ \bibinfo {pages} {939}
  (\bibinfo {year} {2010})}\BibitemShut {NoStop}%
\bibitem [{\citenamefont {Okun}(1989)}]{Okun1989}%
  \BibitemOpen
  \bibfield  {author} {\bibinfo {author} {\bibfnamefont {L.~B.}\ \bibnamefont
  {Okun}},\ }\bibfield  {title} {\bibinfo {title} {Tests of electric charge
  conservation and the {P}auli principle},\ }\href@noop {} {\bibfield
  {journal} {\bibinfo  {journal} {Sov. Phys. Usp.}\ }\textbf {\bibinfo {volume}
  {32}},\ \bibinfo {pages} {543} (\bibinfo {year} {1989})}\BibitemShut
  {NoStop}%
\bibitem [{\citenamefont {Arnquist}\ \emph {et~al.}(2024)\citenamefont
  {Arnquist} \emph {et~al.}}]{ArnquistNatPhys2024}%
  \BibitemOpen
  \bibfield  {author} {\bibinfo {author} {\bibfnamefont {I.~J.}\ \bibnamefont
  {Arnquist}} \emph {et~al.},\ }\bibfield  {title} {\bibinfo {title} {Search
  for charge non-conservation and {P}auli exclusion principle violation with
  the {M}ajorana {D}emonstrator},\ }\href@noop {} {\bibfield  {journal}
  {\bibinfo  {journal} {Nat. Phys.}\ }\textbf {\bibinfo {volume} {20}},\
  \bibinfo {pages} {1078} (\bibinfo {year} {2024})}\BibitemShut {NoStop}%
\bibitem [{\citenamefont {Zhang}\ \emph {et~al.}(2025)\citenamefont {Zhang},
  \citenamefont {Geng}, \citenamefont {Hu},\ and\ \citenamefont
  {Zhang}}]{ZhangPRD2025}%
  \BibitemOpen
  \bibfield  {author} {\bibinfo {author} {\bibfnamefont {Z.~H.}\ \bibnamefont
  {Zhang}}, \bibinfo {author} {\bibfnamefont {X.~P.}\ \bibnamefont {Geng}},
  \bibinfo {author} {\bibfnamefont {J.~W.}\ \bibnamefont {Hu}},\ and\ \bibinfo
  {author} {\bibfnamefont {Z.~H.}\ \bibnamefont {Zhang}},\ }\bibfield  {title}
  {\bibinfo {title} {Test of electric charge conservation in dark matter direct
  detection experiments},\ }\href@noop {} {\bibfield  {journal} {\bibinfo
  {journal} {Phys. Rev. D}\ }\textbf {\bibinfo {volume} {112}},\ \bibinfo
  {pages} {035032} (\bibinfo {year} {2025})}\BibitemShut {NoStop}%
\bibitem [{\citenamefont {Agostini}\ \emph {et~al.}(2015)\citenamefont
  {Agostini} \emph {et~al.}}]{AgostiniPRL2015}%
  \BibitemOpen
  \bibfield  {author} {\bibinfo {author} {\bibfnamefont {M.}~\bibnamefont
  {Agostini}} \emph {et~al.},\ }\bibfield  {title} {\bibinfo {title} {Test of
  electric charge conservation with {B}orexino},\ }\href@noop {} {\bibfield
  {journal} {\bibinfo  {journal} {Phys. Rev. Lett.}\ }\textbf {\bibinfo
  {volume} {115}},\ \bibinfo {pages} {231802} (\bibinfo {year}
  {2015})}\BibitemShut {NoStop}%
\bibitem [{\citenamefont {Weinberg}(1964)}]{Weinberg1964}%
  \BibitemOpen
  \bibfield  {author} {\bibinfo {author} {\bibfnamefont {S.}~\bibnamefont
  {Weinberg}},\ }\bibfield  {title} {\bibinfo {title} {Photons and gravitons in
  {$S$}-matrix theory: derivation of charge conservation and equality of
  gravitational and inertial mass},\ }\href@noop {} {\bibfield  {journal}
  {\bibinfo  {journal} {Phys. Rev.}\ }\textbf {\bibinfo {volume} {135}},\
  \bibinfo {pages} {B1049} (\bibinfo {year} {1964})}\BibitemShut {NoStop}%
\bibitem [{\citenamefont {Berman}(2008)}]{BermanAmJPhys2008}%
  \BibitemOpen
  \bibfield  {author} {\bibinfo {author} {\bibfnamefont {P.~R.}\ \bibnamefont
  {Berman}},\ }\bibfield  {title} {\bibinfo {title} {Dynamic creation of
  electrostatic fields},\ }\href@noop {} {\bibfield  {journal} {\bibinfo
  {journal} {Am. J. Phys.}\ }\textbf {\bibinfo {volume} {76}},\ \bibinfo
  {pages} {48} (\bibinfo {year} {2008})}\BibitemShut {NoStop}%
\bibitem [{\citenamefont {Alicki}(1978)}]{AlickiJPA1978}%
  \BibitemOpen
  \bibfield  {author} {\bibinfo {author} {\bibfnamefont {R.}~\bibnamefont
  {Alicki}},\ }\bibfield  {title} {\bibinfo {title} {Generalised
  electrodynamics},\ }\href@noop {} {\bibfield  {journal} {\bibinfo  {journal}
  {J. Phys. A: Math. Gen.}\ }\textbf {\bibinfo {volume} {11}},\ \bibinfo
  {pages} {1807} (\bibinfo {year} {1978})}\BibitemShut {NoStop}%
\bibitem [{\citenamefont {L\"ammerzahl}\ \emph {et~al.}(2005)\citenamefont
  {L\"ammerzahl}, \citenamefont {Mac\'{\i}as},\ and\ \citenamefont
  {M\"uller}}]{MullerPRD2005}%
  \BibitemOpen
  \bibfield  {author} {\bibinfo {author} {\bibfnamefont {C.}~\bibnamefont
  {L\"ammerzahl}}, \bibinfo {author} {\bibfnamefont {A.}~\bibnamefont
  {Mac\'{\i}as}},\ and\ \bibinfo {author} {\bibfnamefont {H.}~\bibnamefont
  {M\"uller}},\ }\bibfield  {title} {\bibinfo {title} {Lorentz invariance
  violation and charge (non)conservation: A general theoretical frame for
  extensions of the {M}axwell equations},\ }\href@noop {} {\bibfield  {journal}
  {\bibinfo  {journal} {Phys. Rev. D}\ }\textbf {\bibinfo {volume} {71}},\
  \bibinfo {pages} {025007} (\bibinfo {year} {2005})}\BibitemShut {NoStop}%
\bibitem [{\citenamefont {{Greiner}}\ and\ \citenamefont
  {{Reinhardt}}(1996)}]{Greiner}%
  \BibitemOpen
  \bibfield  {author} {\bibinfo {author} {\bibfnamefont {R.}~\bibnamefont
  {{Greiner}}}\ and\ \bibinfo {author} {\bibfnamefont {J.}~\bibnamefont
  {{Reinhardt}}},\ }\href@noop {} {\emph {\bibinfo {title} {{Field
  Quantization}}}}\ (\bibinfo  {publisher} {Springer-Verlag},\ \bibinfo {year}
  {1996})\BibitemShut {NoStop}%
\bibitem [{\citenamefont {Blaschke}\ and\ \citenamefont
  {Gieres}(2021)}]{GieresNPB2021}%
  \BibitemOpen
  \bibfield  {author} {\bibinfo {author} {\bibfnamefont {D.~N.}\ \bibnamefont
  {Blaschke}}\ and\ \bibinfo {author} {\bibfnamefont {F.}~\bibnamefont
  {Gieres}},\ }\bibfield  {title} {\bibinfo {title} {On the canonical
  formulation of gauge field theories and {P}oincaré transformations},\
  }\href@noop {} {\bibfield  {journal} {\bibinfo  {journal} {Nucl. Phys. B}\
  }\textbf {\bibinfo {volume} {965}},\ \bibinfo {pages} {115366} (\bibinfo
  {year} {2021})}\BibitemShut {NoStop}%
\bibitem [{\citenamefont {Minotti}\ and\ \citenamefont
  {Modanese}(2023)}]{ModaneseSym2023}%
  \BibitemOpen
  \bibfield  {author} {\bibinfo {author} {\bibfnamefont {F.}~\bibnamefont
  {Minotti}}\ and\ \bibinfo {author} {\bibfnamefont {G.}~\bibnamefont
  {Modanese}},\ }\bibfield  {title} {\bibinfo {title} {{A}haronov-{B}ohm
  electrodynamics in material media: A scalar e.m. field cannot cause
  dissipation in a medium},\ }\href@noop {} {\bibfield  {journal} {\bibinfo
  {journal} {Symmetry}\ }\textbf {\bibinfo {volume} {15}},\ \bibinfo {pages}
  {1119} (\bibinfo {year} {2023})}\BibitemShut {NoStop}%
\bibitem [{\citenamefont {Beltr\'an~Jim\'enez}\ and\ \citenamefont
  {Maroto}(2011)}]{JimenezPRD2011}%
  \BibitemOpen
  \bibfield  {author} {\bibinfo {author} {\bibfnamefont {J.}~\bibnamefont
  {Beltr\'an~Jim\'enez}}\ and\ \bibinfo {author} {\bibfnamefont {A.~L.}\
  \bibnamefont {Maroto}},\ }\bibfield  {title} {\bibinfo {title} {Cosmological
  magnetic fields from inflation in extended electromagnetism},\ }\href@noop {}
  {\bibfield  {journal} {\bibinfo  {journal} {Phys. Rev. D}\ }\textbf {\bibinfo
  {volume} {83}},\ \bibinfo {pages} {023514} (\bibinfo {year}
  {2011})}\BibitemShut {NoStop}%
\bibitem [{\citenamefont {Goldhaber}\ and\ \citenamefont
  {Nieto}(1971)}]{Nieto_RMP1971}%
  \BibitemOpen
  \bibfield  {author} {\bibinfo {author} {\bibfnamefont {A.~S.}\ \bibnamefont
  {Goldhaber}}\ and\ \bibinfo {author} {\bibfnamefont {M.~M.}\ \bibnamefont
  {Nieto}},\ }\bibfield  {title} {\bibinfo {title} {Terrestrial and
  extraterrestrial limits on the photon mass},\ }\href@noop {} {\bibfield
  {journal} {\bibinfo  {journal} {Rev. Mod. Phys.}\ }\textbf {\bibinfo {volume}
  {43}},\ \bibinfo {pages} {277} (\bibinfo {year} {1971})}\BibitemShut
  {NoStop}%
\bibitem [{\citenamefont {Damski}(2023)}]{BDPeriodic1}%
  \BibitemOpen
  \bibfield  {author} {\bibinfo {author} {\bibfnamefont {B.}~\bibnamefont
  {Damski}},\ }\bibfield  {title} {\bibinfo {title} {Periodic charge
  oscillations in the {P}roca theory},\ }\href@noop {} {\bibfield  {journal}
  {\bibinfo  {journal} {Nucl. Phys. B}\ }\textbf {\bibinfo {volume} {994}},\
  \bibinfo {pages} {116300} (\bibinfo {year} {2023})}\BibitemShut {NoStop}%
\bibitem [{\citenamefont {Damski}(2025)}]{BDCounter}%
  \BibitemOpen
  \bibfield  {author} {\bibinfo {author} {\bibfnamefont {B.}~\bibnamefont
  {Damski}},\ }\bibfield  {title} {\bibinfo {title} {Non-equilibrium dynamics
  of long-range field configurations in the {P}roca theory and the
  counterexample to the law of periodic charge oscillations},\ }\href@noop {}
  {\bibfield  {journal} {\bibinfo  {journal} {Nucl. Phys. B}\ }\textbf
  {\bibinfo {volume} {1020}},\ \bibinfo {pages} {117132} (\bibinfo {year}
  {2025})}\BibitemShut {NoStop}%
\bibitem [{\citenamefont {{Gerry}}\ and\ \citenamefont
  {{Knight}}(2005)}]{Gerry}%
  \BibitemOpen
  \bibfield  {author} {\bibinfo {author} {\bibfnamefont {C.~C.}\ \bibnamefont
  {{Gerry}}}\ and\ \bibinfo {author} {\bibfnamefont {P.~L.}\ \bibnamefont
  {{Knight}}},\ }\href@noop {} {\emph {\bibinfo {title} {{Introductory Quantum
  Optics}}}}\ (\bibinfo  {publisher} {Cambridge University Press},\ \bibinfo
  {year} {2005})\BibitemShut {NoStop}%
\bibitem [{\citenamefont {Scully}\ and\ \citenamefont
  {Zubairy}(1997)}]{Scully}%
  \BibitemOpen
  \bibfield  {author} {\bibinfo {author} {\bibfnamefont {M.~O.}\ \bibnamefont
  {Scully}}\ and\ \bibinfo {author} {\bibfnamefont {M.~S.}\ \bibnamefont
  {Zubairy}},\ }\href@noop {} {\emph {\bibinfo {title} {Quantum Optics}}}\
  (\bibinfo  {publisher} {Cambridge University Press},\ \bibinfo {year}
  {1997})\BibitemShut {NoStop}%
\bibitem [{\citenamefont {Kulish}\ and\ \citenamefont
  {Faddeev}(1970)}]{KulishFaddeev}%
  \BibitemOpen
  \bibfield  {author} {\bibinfo {author} {\bibfnamefont {P.~P.}\ \bibnamefont
  {Kulish}}\ and\ \bibinfo {author} {\bibfnamefont {L.~D.}\ \bibnamefont
  {Faddeev}},\ }\bibfield  {title} {\bibinfo {title} {Asymptotic conditions and
  infrared divergences in quantum electrodynamics},\ }\href@noop {} {\bibfield
  {journal} {\bibinfo  {journal} {Theor. Math. Phys.}\ }\textbf {\bibinfo
  {volume} {4}},\ \bibinfo {pages} {745} (\bibinfo {year} {1970})}\BibitemShut
  {NoStop}%
\bibitem [{\citenamefont {Furugori}\ and\ \citenamefont
  {Nojiri}(2021)}]{FurugoriPRD2021}%
  \BibitemOpen
  \bibfield  {author} {\bibinfo {author} {\bibfnamefont {H.}~\bibnamefont
  {Furugori}}\ and\ \bibinfo {author} {\bibfnamefont {S.}~\bibnamefont
  {Nojiri}},\ }\bibfield  {title} {\bibinfo {title} {{Dressed asymptotic states
  and QED infrared physics}},\ }\href@noop {} {\bibfield  {journal} {\bibinfo
  {journal} {Phys. Rev. D}\ }\textbf {\bibinfo {volume} {104}},\ \bibinfo
  {pages} {125004} (\bibinfo {year} {2021})}\BibitemShut {NoStop}%
\bibitem [{\citenamefont {{Gradshteyn}}\ \emph {et~al.}(2014)\citenamefont
  {{Gradshteyn}}, \citenamefont {{Ryzhik}}, \citenamefont {{Zwillinger}},\ and\
  \citenamefont {{Moll}}}]{Ryzhik}%
  \BibitemOpen
  \bibfield  {author} {\bibinfo {author} {\bibfnamefont {I.~S.}\ \bibnamefont
  {{Gradshteyn}}}, \bibinfo {author} {\bibfnamefont {I.~M.}\ \bibnamefont
  {{Ryzhik}}}, \bibinfo {author} {\bibfnamefont {D.}~\bibnamefont
  {{Zwillinger}}},\ and\ \bibinfo {author} {\bibfnamefont {V.}~\bibnamefont
  {{Moll}}},\ }\href@noop {} {\emph {\bibinfo {title} {{Table of Integrals,
  Series, and Products}}}},\ \bibinfo {edition} {8th}\ ed.\ (\bibinfo
  {publisher} {Academic Press},\ \bibinfo {year} {2014})\BibitemShut {NoStop}%
\bibitem [{Mat()}]{Mathematica143}%
  \BibitemOpen
  \href@noop {} {}\bibinfo {note} {Wolfram Research, Inc., Mathematica, Version
  14.3, Champaign, IL (2025).}\BibitemShut {Stop}%
\bibitem [{\citenamefont {Rohrlich}(2007)}]{RohrlichClassicalBook}%
  \BibitemOpen
  \bibfield  {author} {\bibinfo {author} {\bibfnamefont {F.}~\bibnamefont
  {Rohrlich}},\ }\href@noop {} {\emph {\bibinfo {title} {Classical Charged
  Particles}}},\ \bibinfo {edition} {3rd}\ ed.\ (\bibinfo  {publisher} {World
  Scientific},\ \bibinfo {year} {2007})\BibitemShut {NoStop}%
\bibitem [{\citenamefont {{Gupta}}(1977)}]{Gupta}%
  \BibitemOpen
  \bibfield  {author} {\bibinfo {author} {\bibfnamefont {S.~N.}\ \bibnamefont
  {{Gupta}}},\ }\href@noop {} {\emph {\bibinfo {title} {{Quantum
  Electrodynamics}}}}\ (\bibinfo  {publisher} {Gordon and Breach Science
  Publishers},\ \bibinfo {year} {1977})\BibitemShut {NoStop}%
\end{thebibliography}

%

\end{document}